\documentclass[12pt]{article}
\usepackage{amsmath, amsthm, amssymb,graphicx}

\pdfoutput=1

\usepackage{latexStyles/sectsty}
\subsectionfont{\it}

\usepackage{hyperref}

\usepackage{latexStyles/apacite}

\renewcommand{\APACjournalVolNumPages}[4]{%
  \Bem{#1}
  \ifx\@empty#2\@empty
  \else
    \unskip\ \textbf{#2:}
  \fi
  \ifx\@empty#3\@empty
  \else
    \unskip
  \fi
  \ifx\@empty#4\@empty
  \else
    \unskip\ {#4}
  \fi
}
\renewcommand{\APACaddressPublisher}[2]{%
  \ifx\@empty#1\@empty
    \ifx\@empty#2\@empty
    \else
      {#1}
    \fi
  \else
    {#2}
    \ifx\@empty#2\@empty
    \else
      \unskip, {#1}
    \fi
  \fi
}

\def\L{\Lambda}
\def\dy{\dd y}
\def\dx{\dd x}
\def\ds{\dd s}
\def\ang#1{\langle#1\rangle}
\def\levy{L{\'e}vy}
\def\F{{\cal F}}
\def\Finv{{\cal F}^{-1}}
\def\Fhat{\hat{F}}

\def\a{\alpha}

\def\d{\delta}

\def\e{\epsilon}
\def\g{\gamma}
\def\k{\kappa}
\def\l{\lambda}
\def\m{\mu}
\def\om{\omega}

\def\s{\sigma}

\def\th{\theta}
\def\bar#1{\overline{#1}}
\def\ovr#1#2{{{#1}\over{#2}}}

\def\dd{{\hbox{\rm d}}}

\def\part{\partial}
\def\povr#1#2{\ovr{\part #1}{\part #2}}

\def\Eq#1{Eq.~\eqref{#1}}
\def\biblist{\begingroup\advance\leftskip by 25pt 
	\parindent-25pt\frenchspacing\noindent}

\def\boldrule{\hrule height 1.2pt}
\def\noterule{\medskip\boldrule\medskip}	

\title{The Common Patterns of Nature}
\author{Steven A.\ Frank\footnote{Department of Ecology and Evolutionary Biology, University of California, Irvine, CA 92697--2525, USA and Santa Fe Institute, 1399 Hyde Park Road, Santa Fe, NM 87501, USA, email: safrank@uci.edu}}
\begin{document}

\maketitle

\vfill
\noterule
\noindent{\bf Please cite as follows:} Frank, S. A. 2009. The common patterns of nature. Journal of Evolutionary Biology XX:XXX--XXX. [Check link below for final volume and page numbers.]
\noterule

\centerline{\bf The published, definitive version of this article is freely available at:} 
\medskip\centerline{\href{http://dx.doi.org/10.1111/j.1420-9101.2009.01775.x}{http://dx.doi.org/10.1111/j.1420-9101.2009.01775.x}}
\noterule

\vfill\eject

\section*{Abstract}

We typically observe large-scale outcomes that arise from the interactions of many hidden, small-scale processes.  Examples include age of disease onset, rates of amino acid substitutions, and composition of ecological communities.   The macroscopic patterns in each problem often vary around a characteristic shape that can be generated by neutral processes.  A neutral generative model assumes that each microscopic process follows unbiased or random stochastic fluctuations: random connections of network nodes; amino acid substitutions with no effect on fitness; species that arise or disappear from communities randomly.  These neutral generative models often match common patterns of nature.  In this paper, I present the theoretical background by which we can understand why these neutral generative models are so successful.  I show where the classic patterns come from, such as the Poisson pattern, the normal or Gaussian pattern, and many others.  Each classic pattern was often discovered by a simple neutral generative model.  The neutral patterns share a special characteristic: they describe the patterns of nature that follow from simple constraints on information.  For example, any aggregation of processes that preserves information only about the mean and variance attracts to the Gaussian pattern; any aggregation that preserves information only about the mean attracts to the exponential pattern; any aggregation that preserves information only about the geometric mean attracts to the power law pattern.  I present a simple and consistent informational framework of the common patterns of nature based on the method of maximum entropy.  This framework shows that each neutral generative model is a special case that helps to discover a particular set of informational constraints; those informational constraints define a much wider domain of non-neutral generative processes that attract to the same neutral pattern.  

\vfill\eject

\begin{quotation}
In fact, all epistemologic value of the theory of probability is based on this: that large-scale random phenomena in their collective action create strict, nonrandom regularity \cite[p.~1]{Gnedenko68Variables}. 
\end{quotation}

\begin{quotation}
I know of scarcely anything so apt to impress the imagination as the wonderful form of cosmic order expressed by the ``law of frequency of error'' [the normal or Gaussian distribution]. Whenever a large sample of chaotic elements is taken in hand and marshaled in the order of their magnitude, this unexpected and most beautiful form of regularity proves to have been latent all along. The law $\ldots$ reigns with serenity and complete self-effacement amidst the wildest confusion. The larger the mob and the greater the apparent anarchy, the more perfect is its sway. It is the supreme law of unreason \cite[p.~166]{Galton89Inheritance}. 
\end{quotation}

\begin{quotation}
We cannot understand what is happening until we learn to think of probability distributions in terms of their demonstrable {\it information content} $\ldots$ \cite[p.~198]{Jaynes03Science}. 
\end{quotation}

\section*{Introduction}

Most patterns in biology arise from aggregation of many small processes.  Variations in the dynamics of complex neural and biochemical networks depend on numerous fluctuations in connectivity and flow through small-scale subcomponents of the network.  Variations in cancer onset arise from variable failures in the many individual checks and balances on DNA repair, cell cycle control, and tissue homeostasis.  Variations in the ecological distribution of species follow the myriad local differences in the birth and death rates of species and in the small-scale interactions between particular species.

In all such complex systems, we wish to understand how large-scale pattern arises from the aggregation of small-scale processes.  A single dominant principle sets the major axis from which all explanation of aggregation and scale must be developed.  This dominant principle is the limiting distribution.

The best known of the limiting distributions, the Gaussian (normal) distribution, follows from the central limit theorem.  If an outcome, such as height, weight, or yield, arises from the summing up of many small-scale processes, then the distribution typically approaches the Gaussian curve in the limit of aggregation over many processes.  

The individual, small-scale fluctuations caused by each contributing process rarely follow the Gaussian curve.  But, with aggregation of many partly uncorrelated fluctuations, each small in scale relative to the aggregate, the sum of the fluctuations smooths into the Gaussian curve---the limiting distribution in this case.  One might say that the numerous small fluctuations tend to cancel out, revealing the particular form of regularity or information that characterizes aggregation and scale for the process under study.

The central limit theorem is widely known, and the Gaussian distribution is widely recognized as an aggregate pattern.  This limiting distribution is so important that one could hardly begin to understand patterns of nature without an instinctive recognition of the relation between aggregation and the Gaussian curve.

In this paper, I discuss biological patterns within a broader framework of limiting distributions.  I emphasize that the common patterns of nature arise from distinctive limiting distributions. In each case, one must understand the distinctive limiting distribution in order to analyze pattern and process.  

In regard to the different limiting distributions that characterize patterns of nature, aggregation and scale have at least three important consequences.  First, a departure from the expected limiting distribution suggests an interesting process that perturbs the typical regularity of aggregation. Such departures from expectation can only be recognized and understood if one has a clear grasp of the characteristic limiting distribution for the particular problem.  

Second, one must distinguish clearly between two distinct meanings of ``neutrality.''  For example, if we count the number of events of a purely random, or ``neutral,'' process, we observe a Poisson pattern.  However, the Poisson may also arise as a limiting distribution by aggregation of small-scale, nonrandom processes.  So we must distinguish between two alternative causes of neutral pattern: the generation of pattern by neutral processes, or the generation of pattern by aggregation of non-neutral processes in which the non-neutral fluctuations tend to cancel in the aggregate.  This distinction cuts to the heart of how we may test neutral theories in ecology and evolution.

Third, a powerfully attracting limiting distribution may be relatively insensitive to perturbations.  Insensitivity arises because, to be attracting over broad variations in the aggregated processes, the limiting distribution must not change too much in response to perturbations.  Insensitivity to perturbation is a reasonable definition of robustness.  Thus, robust patterns may often coincide with limiting distributions. 

In general, inference in biology depends critically on understanding the nature of limiting distributions.  If a pattern can only be generated by a very particular hypothesized process, then observing the pattern strongly suggests that the pattern was created by the hypothesized generative process.  By contrast, if the same pattern arises as a limiting distribution from a variety of underlying processes, then a match between theory and pattern only restricts the underlying generative processes to the broad set that attracts to the limiting pattern.  Inference must always be discussed in relation to the breadth of processes attracted to a particular pattern. Because many patterns in nature arise from limiting distributions, such distributions form the core of inference with regard to the relations between pattern and process.

I recently took up study of the problems outlined in this introduction.  In my studies, I found it useful to separate the basic facts of probability theory that set the background from my particular ideas about biological robustness, the neutral theories in ecology and evolution, and the causes of patterns such as the age of cancer onset and the age of death.  The basic facts of probability are relatively uncontroversial, whereas my own interpretations of particular biological patterns remain open to revision and to the challenge of empirical tests.  

In this paper, I focus on the basic facts of probability theory to set the background for future work.  Thus, this paper serves primarily as a tutorial to aspects of probability framed in the context of several recent conceptual advances from the mathematical and physical sciences.   In particular, I use Jaynes' maximum entropy approach to unify the relations between aggregation and pattern  \cite{Jaynes03Science}.  Information plays the key role.  In each problem, ultimate pattern arises from the particular information preserved in the face of the combined fluctuations in aggregates that decay all non-preserved aspects of pattern toward maximum entropy or maximum randomness.  My novel contribution is to apply this framework of entropy and information in a very simple and consistent way across the full range of common patterns in nature. 

\section*{Overview}

The general issues of aggregation and limiting distributions are widely known throughout the sciences, particularly in physics.  These concepts form the basis for much of statistical mechanics and thermodynamics. Yet, in spite of all of this work, the majority of research in biology remains outside the scope of these fundamental and well understood principles.  Indeed, much of the biological literature continues as if the principles of aggregation and scale hardly existed, in spite of a few well argued publications that appear within each particular subject.  

Three reasons may explain the disconnect between, on the one hand, the principles of aggregation and scale, and, on the other hand, the way in which different subfields of biology often deal with the relations between pattern and process.  First, biologists are often unaware of the general quantitative principles, and lack experience with and exposure to other quantitative fields of science.  Second, each subfield within biology tends to struggle with the fundamental issues independently of other subfields.  Little sharing occurs of the lessons slowly learned within distinct subfields. Third, in spite of much mathematical work on aggregation and limiting distributions, many fundamental issues remain unresolved or poorly connected to commonly observed phenomena.  For example, power law distributions arise in economics and in nearly every facet of nature. Almost all theories to explain the observed power law patterns emphasize either particular models of process, which have limited scope, or overly complex theories of entropy and aggregation, which seem too specialized to form a general explanation.  

I begin with two distinct summaries of the basic quantitative principles.  In the first summary, I discuss in an intuitive way some common distributions that describe patterns of nature: the exponential, Poisson, and gamma distributions.  These distributions often associate with the concepts of random or neutral processes, setting default expectations about pattern.  I emphasize the reason these patterns arise so often:  combinations of many small-scale processes tend to yield, at a higher, aggregate scale, these common distributions.  We call the observed patterns from such aggregation the limiting distributions, because these distributions arise in the limit of aggregation, at a larger scale, over many partially uncorrelated processes at smaller scales.  The small-scale fluctuations tend to cancel out at the larger scale, yielding observed patterns that appear random, or neutral, even though the pattern is created by aggregation of nonrandom processes.  

In my second summary of the basic quantitative principles, I give a slightly more technical description.  In particular, I discuss what ``random'' means, and why aggregation of nonrandom processes leads in the limit to the random or neutral patterns at larger scales.  

The following sections discuss aggregation more explicitly.  I emphasize how processes interact to form aggregate pattern, and how widespread such aggregation must be.  Certain general patterns arise as the most fundamental outcomes of aggregation, of which the Gaussian distribution and the central limit theorem are special cases.

Finally, I conclude that the most important consequences of aggregation are simple yet essential to understand: certain patterns dominate for particular fields of study; those patterns set a default against which deviations must be understood; and the key to inference turns on understanding the separation between those small-scale processes that tend to attract in the aggregate to the default pattern and those small-scale processes that do not aggregate to the default pattern.

\section*{Common distributions}

This section provides brief, intuitive descriptions of a few important patterns expressed as probability distributions. These descriptions set the stage for the more comprehensive presentation in the following section.  Numerous books provide introductions to common probability distributions and relevant aspects of probability theory \shortcite<e.g., >{Feller68Applications,Feller71Applications,Johnson94Distributions,Johnson05Distributions,Kleiber03Sciences}.

\subsection*{Poisson}

One often counts the number of times an event occurs per unit area or per unit time.  When the 
numbers per unit of measurement are small, the observations often follow the Poisson distribution.  The Poisson defines the neutral or random expectation for pattern, because the Poisson pattern arises when events are scattered at random over a grid of spatial or temporal units.  

A standard test for departure from randomness compares observations against the random expectations that arise from the Poisson distribution.  A common interpretation from this test is that a match between observations and the Poisson expectation implies that the pattern was generated by a neutral or random process.  However, to evaluate how much information one obtains from such a match, one must know how often aggregations of nonrandom processes may generate the same random pattern.

Various theorems of probability tell us when particular combinations of underlying, smaller scale processes lead to a Poisson pattern at the larger, aggregate scale.  Those theorems, which I discuss in a later section, define the ``law of small numbers.''  From those theorems, we get a sense of the basin of attraction to the Poisson as a limiting distribution: in other words, we learn about the separation between those aggregations of small-scale processes that combine to attract toward the Poisson pattern and those aggregations that do not.  

Here is a rough, intuitive way to think about the basin of attraction toward a random, limiting distribution \cite{Jaynes03Science}.  Think of each small-scale process as contributing a deviation from random in the aggregate.  If many nonrandom and partially uncorrelated small-scale deviations aggregate to form an overall pattern, then the individual nonrandom deviations will often cancel in the aggregate and attract to the limiting random distribution.  Such smoothing of small-scale deviations in the aggregate pattern must be rather common, explaining why a random pattern with strict regularity, such as the Poisson, is so widely observed.

Many aggregate patterns will, of course, deviate from the random limiting distribution.  To understand such deviations, we must consider them in relation to the basin of attraction for random pattern.  In general, the basins of attraction of random patterns form the foundation by which we must interpret observations of nature.  In this regard, the major limiting distributions are the cornerstone of natural history.

\subsection*{Exponential and gamma}

The waiting time for the first occurrence of some particular event often follows an exponential distribution.  The exponential has three properties that associate it with a neutral or random pattern.  

First, the waiting time until the first occurrence of a Poisson process has an exponential distribution: the exponential is the time characterization of how long one waits for random events, and the Poisson is the count characterization of the number of random events that occur in a fixed time (or space) interval.  

Second, the exponential distribution has the memoryless property.  Suppose the waiting time for the first occurrence of an event follows the exponential distribution.  If, starting at time $t=0$, the event does not occur during the interval up to time $t=T$, then the waiting time for the first occurrence starting from time $T$ is the same as it was when we began to measure from time zero.  In other words, the process has no memory of how long it has been waiting; occurrences happen at random irrespective of history.  

Third, the exponential is in many cases the limiting distribution for aggregation of smaller-scale waiting time processes (see below).  For example, time to failure patterns often follow the exponential, because the failure of a machine or aggregate biological structure often depends on the time to failure of any essential component.  Each component may have a time to failure that differs from exponential, but in the aggregate, the waiting time for the overall failure of the machine often converges to the exponential.  

The gamma distribution arises in many ways, a characteristic of limiting distributions.  For example, if the waiting time until the first occurrence of a random process follows an exponential, and occurrences happen independently of each other, then the waiting time for the $n$th occurrence follows the gamma pattern.  The exponential is a limiting distribution with a broad basin of attraction.  Thus, the gamma is also a limiting distribution that attracts many aggregate patterns. For example, if an aggregate structure fails only after multiple subcomponents fail, then the time to failure may in the limit attract to a gamma pattern.  In a later section, I will discuss a more general way in which to view the gamma distribution with respect to randomness and limiting distributions.

\subsection*{Power law}

Many patterns of nature follow a power law distribution \cite{Mandelbrot83Nature,Kleiber03Sciences,Mitzenmacher04distributions,Newman05law,Simkin06Willis,Sornette06Tools}.  Consider the distribution of wealth in human populations as an example.  Suppose that the frequency of individuals with wealth $x$ is $f(x)$, and the frequency  with twice that wealth is $f(2x)$.  Then the ratio of those with wealth $x$ relative to those with twice that wealth is $f(x)/f(2x)$.  That ratio of wealth is often constant no matter what level of baseline wealth, $x$, that we start with, so long as we look above some minimum value of wealth, $L$.  In particular,
	$$
		\ovr{f(x)}{f(2x)}=k,
	$$
where $k$ is a constant, and $x>L$.   Such relations are called ``scale invariant,'' because no matter how big or small $x$, that is, no matter what scale at which we look, the change in frequency follows the same constant pattern.  

Scale-invariant pattern implies a power law relationship for the frequencies
	$$f(x) = ax^{-b},$$
where $a$ is an uninteresting constant that must be chosen so that the total frequency sums to one, and $b$ is a constant that sets how fast wealth becomes less frequent as wealth increases.  For example, a doubling in wealth leads to 
	$$\ovr{f(x)}{f(2x)}=\ovr{ax^{-b}}{a(2x)^{-b}}=2^b,$$
which shows that the ratio of the frequency of people with wealth $x$ relative to those with wealth $2x$ does not depend on the initial wealth, $x$, that is, it does not depend on the scale at which we look.

Scale invariance, expressed by power laws, describes a very wide range of natural patterns.  To give just a short listing, \citeA{Sornette06Tools} mentions that the following phenomena follow power law distributions: the magnitudes of earthquakes, hurricanes, volcanic eruptions, and floods; the sizes of meteorites; and losses caused by business interruptions from accidents.  Other studies have documented power laws in stock market fluctuations, sizes of computer files, and word frequency in languages \cite{Mitzenmacher04distributions,Newman05law,Simkin06Willis}.  

In biology, power laws have been particularly important in analyzing connectivity patterns in metabolic networks \shortcite{Barabasi99networks,Ravasz02networks} and in the number of species observed per unit area in ecology \cite{Garcia06ecology}.

Many models have been developed to explain why power laws arise.  Here is a simple example from \citeA{Simon55functions} to explain the power law distribution of word frequency in languages \cite<see>{Simkin06Willis}.  Suppose we start with a collection of $N$ words.  We then add another word.  With probability $p$, the word is new.  With probability $1-p$, the word matches one already in our collection; the particular match to an existing word occurs with probability proportional to the relative frequencies of existing words.  In the long run, the frequency of words that occurs $x$ times is proportional to $x^{-[1+1/(1-p)]}$. We can think of this process as preferential attachment, or an example in which the rich get richer.  

Simon's model sets out a simple process that generates a power law and fits the data.  But could other  simple processes generate the same pattern?  We can express this question in an alternative way, following the theme of this paper: What is the basin of attraction for processes that converge onto the same pattern?  The following sections take up this question and, more generally, how we may think about the relationship between generative models of process and the commonly observed patterns that result.

\section*{Random or neutral distributions}

Much of biological research is reverse engineering.  We observe a pattern or design, and we try to infer the underlying process that generated what we see.  The observed patterns can often be described as probability distributions: the frequencies of genotypes; the numbers of nucleotide substitutions per site over some stretch of DNA; the different output response strengths or movement directions given some input; or the numbers of species per unit area.

The same small set of probability distributions describe the great majority of observed patterns: the binomial, Poisson, Gaussian, exponential, power law, gamma, and a few other common distributions.  These distributions reveal the contours of nature.  We must understand why these distributions are so common and what they tell us, because our goal is to use these observed patterns to reverse engineer the underlying processes that created those patterns.  What information do these distributions contain?  

\subsection*{Maximum entropy}

The key probability distributions often arise as the most random pattern consistent the information expressed by a few constraints \cite{Jaynes03Science}.  In this section, I introduce the concept of maximum entropy, where entropy measures randomness. In the following sections, I derive common distributions to show how they arise from maximum entropy (randomness) subject to constraints such as information about the mean, variance, or geometric mean.  My mathematical presentation throughout is informal and meant to convey basic concepts. Those readers interested in the mathematics should follow the references to the original literature.

The probability distributions follow from Shannon's measure of information \cite{Shannon49Communication}. I first define this measure of information. I then discuss an intuitive way of thinking about the measure and its relation to entropy.

Consider a probability distribution function (pdf) defined as $p(y|\th)$.  Here, $p$ is the probability of some measurement $y$ given a set of parameters, $\th$. Let the abbreviation $p_y$ stand for $p(y|\th)$.  Then Shannon information is defined as
	$$H = - \sum p_y\log(p_y),$$
where the sum is taken over all possible values of $y$, and the $\log$ is taken as the natural logarithm.

The value $-\log(p_y)=\log(1/p_y)$ rises as the probability $p_y$ of observing a particular value of $y$ becomes smaller.  In other words,  $-\log(p_y)$ measures the surprise in observing a particular value of $y$, because rare events are more surprising \cite{Tribus61Applications}.  Greater surprise provides more information: if we are surprised by an observation, we learn a lot; if we are not surprised, we had already predicted the outcome to be likely, and we gain little information.  With this interpretation, Shannon information, $H$, is simply an average measure of surprise over all possible values of $y$.   

We may interpret the maximum of $H$ as the least predictable and most random distribution within the constraints of any particular problem.  In physics, randomness is usually expressed in terms of entropy, or disorder, and is measured by the same expression as Shannon information.  Thus, the technique of maximizing $H$ to obtain the most random distribution subject to the constraints of a particular problem is usually referred to as the method of maximum entropy \cite{Jaynes03Science}.

Why should observed probability distributions tend toward those with maximum entropy?  Because observed patterns typically arise by aggregation of many small scale processes.  Any directionality or nonrandomness caused by each small scale process tends, on average, to be canceled in the aggregate: one fluctuation pushes in one direction, another fluctuation pushes in a different direction, and so on.  Of course, not all observations are completely random.  The key is that each problem typically has a few constraints that set the pattern in the aggregate, and all other fluctuations cancel as the nonconstrained aspects tend to the greatest entropy or randomness.  In terms of information, the final pattern reflects only the information content of the system expressed by the constraints on randomness; all else dissipates to maximum entropy as the pattern converges to its limiting distribution defined by its informational constraints \cite{Van81probability}.  

\subsection*{The discrete uniform distribution}

We can find the most probable distribution for a particular problem by the method of maximum entropy.  We simply solve for the probability distribution that maximizes entropy subject to the constraint that the distribution must satisfy any measurable information that we can obtain about the distribution or any assumption that we make about the distribution.

Consider the simplest problem, in which we know that $y$ falls within some bounds $a\le y\le b$, and we require that the total probability sums to one, $\sum_y p_y=1$.  We must also specify what values $y$ may take on between $a$ and $b$.  In the first case, restrict $y$ to the values of integers, so that $y= a,a+1, a+2,\ldots,b$, and there are $N=b-a+1$ possible values for $y$.

We find the maximum entropy distribution by maximizing Shannon entropy, $H$, subject to the constraint that the total probability sums to one, $\sum_y p_y=1$.  We can abbreviate this constraint as $P=\sum_y p_y-1$. By the method of Lagrangian multipliers, this yields the quantity to be maximized as
	$$\L = H - \psi P = -\sum_y p_y\log(p_y)-\psi\Big(\sum_y p_y-1\Big).$$
We have to choose each $p_y$ so that the set maximizes $\L$.  We find that set by solving for each $p_y$ as the value at which the derivative of $\L$ with respect to $p_y$ is zero
	$$\povr{\L}{p_y}=-1-\log(p_y)-\psi=0.$$
Solving yields 
	$$p_y=e^{-(1+\psi)}.$$
To complete the solution, we must find the value of $\psi$, which we can obtain by using the information that the sum over all probabilities is one, thus 
	$$\sum_{y=a}^b p_y = \sum_{y=a}^b e^{-(1+\psi)}=Ne^{-(1+\psi)}=1,$$
where $N$ arises because $y$ takes on $N$ different values ranging from $a$ to $b$.  From this equation, $e^{-(1+\psi)}=1/N$, yielding the uniform distribution
	$$p_y=1/N$$
for $y=a,\ldots,b$.  This result simply says that if we do not know anything except the possible values of our observations and the fact that the total probability is one, then we should consider all possible outcomes equally (uniformly) probable.  The uniform distribution is sometimes discussed as the expression of ignorance or lack of information.  

In observations of nature, we usually can obtain some additional information from measurement or from knowledge about the structure of the problem.  Thus, the uniform distribution does not describe well many patterns of nature, but rather arises most often as an expression of prior ignorance before we obtain information. 

\subsection*{The continuous uniform distribution}

The previous section derived the uniform distribution in which the observations $y$ take on integer values $a, a+1, a+2,\ldots, b$.   In this section, I show the steps and notation for the continuous uniform case.  See \citeA{Jaynes03Science} for technical issues that may arise when analyzing maximum entropy for continuous variables.

Everything is the same as the previous section, except that $y$ can take on any continuous value between $a$ and $b$.  We can move from the discrete case to the continuous case by writing the possible values of $y$ as $a, a+\dy, a+2\dy,\ldots,b$.  In the discrete case above, $\dy=1$.  In the continuous case, we let $\dy\rightarrow0$, that is, we let $\dy$ become arbitrarily small.  Then the number of steps between $a$ and $b$ is $(b-a)/\dy$.

The analysis is exactly as above, but each increment must be weighted by $\dy$, and instead of writing
	$$\sum_{y=a}^b p_y\dy =1$$
we write 
	$$\int_a^bp_y\dy =1$$
to express integration of small units, $\dy$, rather than summation of discrete units.  Then, repeating the key equations from above in the continuous notation, we have the basic expression of the value to be maximized as
	$$\L = H - \psi P = -\int_y p_y\log(p_y)\dy-\psi\bigg(\int_y p_y\dy-1\bigg).$$
From the prior section, we know $\part\L/\part p_y=0$ leads to $p_y= e^{-(1+\psi)}$, thus
	$$\int_a^b p_y\dy = \int_a^b e^{-(1+\psi)}\dy=(b-a)e^{-(1+\psi)}=1,$$
where $b-a$ arises because $\int_a^b\dy=b-a$, thus $e^{-(1+\psi)}=1/(b-a)$, and
	$$p_y = \ovr{1}{b-a},$$
which is the uniform distribution over the continuous interval between $a$ and $b$.

\subsection*{The binomial distribution}

The binomial distribution describes the outcome of a series of $i=1,2,\ldots,N$ observations or trials. Each observation can take on one of two values, $x_i=0$ or $x_i=1$, for the $i$th observation.  For convenience, we refer to an observation of one as a success, and an observation of zero as a failure. We assume each observation is independent of the others, and the probability of a success on any trial is $a_i$, where $a_i$ may vary from trial to trial.  The total number of successes over $N$ trials is $y=\sum x_i$.  

Suppose this is all the information that we have.  We know that our random variable, $y$, can take on a series of integer values, $y=0,1,\ldots,N$, because we may have between zero and $N$ total successes in $N$ trials.  Define the probability distribution as $p_y$, the probability that we observe $y$ successes in $N$ trials. We know that the probabilities sum to one.  Given only that information, it may seem, at first glance, that the maximum entropy distribution would be uniform over the possible outcomes for $y$.  However, the structure of the problem provides more information, which we must incorporate.

How many different ways can we can obtain $y=0$ successes in $N$ trials? Just one: a series of failures on every trial.  How many different ways can we obtain $y=1$ success? There are $N$ different ways: a success on the first trial and failures on the others; a success on the second trial, and failures on the others; and so on.  

The uniform solution by maximum entropy tells us that each different combination is equally likely.  Because each value of $y$ maps to a different number of combinations, we must make a correction for the fact that measurements on $y$ are distinct from measurements on the equally likely combinations.  In particular, we must formulate a measure, $m_y$, that accounts for how the uniformly distributed basis of combinations translates into variable values of the number of successes, $y$.  Put another way, $y$ is invariant to changes in the order of outcomes given a fixed number of successes. That invariance captures a lack of information that must be included in our analysis.

This use of a transform to capture the nature of measurement in a particular problem recurs in analyses of entropy.  The proper analysis of entropy must be made with respect to the underlying measure.  We replace the Shannon entropy with the more general expression
\begin{equation}\label{KLentropy}
	S=-\sum p_y \log\left[\ovr{p_y}{m_y}\right], 
\end{equation}
a measure of relative entropy that is related to the Kullback-Leibler divergence \cite{Kullback59Statistics}.  When $m_y$ is a constant, expressing a uniform transformation, then we recover the standard expression for Shannon entropy.  

In the binomial sampling problem, the number of combinations for each value of $y$ is 
\begin{equation}\label{binomialMeasure}
	m_y = {N\choose y}=\ovr{N!}{y!(N-y)!}. 
\end{equation}
	
Suppose that we also know the expected number of successes in a series of $N$ trials, given as $\ang{y}=\sum yp_y$, where I use the physicists' convention of angle brackets for the expectation of the quantity inside.  Earlier, I defined $a_i$ as the probability of success in the $i$th trial. Note that the average probability of success per trial is $\ang{y}/N=\ang{a}$.  For convenience, let $\a=\ang{a}$, thus the expected number of successes is $\ang{y}=N\a$.

What is the maximum entropy distribution given all of the information that we have, including the expected number of successes?  We proceed by maximizing $S$ subject to the constraint that all the probabilities must add to one and subject to the constraint, $C_1=\sum yp_y-N\a$, that the mean number of successes must be $\ang{y}=N\a$.  The quantity to maximize is
\begin{align}\notag
		\L&=S-\psi P-\l_1C_1\notag\\
			&=-\sum p_y \log\left[\ovr{p_y}{m_y}\right]-\psi\left(\sum p_y-1\right)
			-\l_1\left(\sum yp_y-N\a\right).\notag
\end{align}
Differentiating with respect to $p_y$ and setting to zero yields
\begin{equation}\label{binomialIntermediate}
	p_y = k{N\choose y}e^{-\l_1 y}, 
\end{equation}
where $k=e^{-(1+\psi)}$, in which $k$ and $\l_1$ are two constants that must be chosen to satisfy the two constraints $\sum p_y=1$ and $\sum yp_y=N\a$. The constants $k=(1-\a)^N$ and $e^{-\l_1}=\a/(1-\a)$ satisfy the two constraints \cite[pp.~115--120]{Sivia06Tutorial} and yield the binomial distribution
	$$p_y = {N\choose y}\a^y(1-\a)^{N-y}.$$

Here is the important conclusion. If all of the information available in measurement reduces to knowledge that we are observing the outcome of a series of binary trials and to knowledge of the average number of successes in $N$ trials, then the observations will follow the binomial distribution.  

In the classic sampling theory approach to deriving distributions, one generates a binomial distribution by a series of independent, identical binary trials in which the probability of success per trial does not vary between trials.  That generative neutral model does create a binomial process---it is a sufficient condition.  

However, many distinct processes may also converge to the binomial pattern.  One only requires  information about the trial-based sampling structure and about the expected number of successes over all trials.  The probability of success may vary between trials \cite{Yu08distribution}. 

Distinct aggregations of small scale processes may smooth to reveal only those two aspects of information---sampling structure and average number of successes---with other measurable forms of information canceling in the aggregate.  Thus, the truly fundamental nature of the binomial pattern arises not from the neutral generative model of identical, independent binary trials, but from the measurable information in observations.  I discuss some additional processes that converge to the binomial in a later section on limiting distributions.

\subsection*{The Poisson distribution}

One often observes a Poisson distribution when counting the number of observations per unit time or per unit area.  The Poisson occurs so often because it arises from the ``law of small numbers,'' in which aggregation of various processes converges to the Poisson when the number of counts per unit is small.  Here, I derive the Poisson as a maximum entropy distribution subject to a constraint on the sampling process and a constraint on the mean number of counts per unit \cite[pp.~121]{Sivia06Tutorial}.

Suppose the unit of measure, such as time or space, is divided into a great number, $N$, of very small intervals.  For whatever item or event we are counting, each interval contains a count of either zero or one that is independent of the counts in other intervals.  This subdivision leads to a binomial process.  The measure for the number of different ways a total count of $y=0,1,\ldots,N$ can arise in the $N$ subdivisions is given by $m_y$ of \Eq{binomialMeasure}.  With large $N$, we can express this measure by using Stirling's approximation
	$$N! \approx \sqrt{2\pi N}(N/e)^N,$$
where $e$ is the base for the natural logarithm.  Using this approximation for large $N$, we obtain
	$$m_y ={N\choose y}= \ovr{N!}{y!(N-y)!}=\ovr{N^y}{y!}.$$ 
Entropy maximization yields \Eq{binomialIntermediate}, in which we can use the large $N$ approximation for $m_y$ to yield
	$$p_y = k\ovr{x^y}{y!},$$
in which $x=Ne^{-\l_1}$. From this equation, the constraint $\sum p_y=1$ leads to the identity $\sum_y x^y/y!=e^x$, which implies $k=e^{-x}$.  The constraint $\sum yp_y=\ang{y}=\m$ leads to the identity $\sum_y yx^y/y!=xe^x$, which implies $x=\m$. These substitutions for $k$ and $x$ yield the Poisson distribution
	$$p_y = \m^y\ovr{e^{-\m}}{y!}.$$

\subsection*{The general solution}

All maximum entropy problems have the same form.  We first evaluate our information about the scale of observations and the sampling scheme. We use this information to determine the measure $m_y$ in the general expression for relative entropy in \Eq{KLentropy}.  We then set $n$ additional constraints that capture all of the available information, each constraint expressed as $C_i = \sum_y f_i(y) p_y - \ang{f_i(y)}$, where the angle brackets denote the expected value.  If the problem is continuous, we use integration as the continuous limit of the summation.  

We always use $P=\sum_y  p_y - 1$ to constrain the total probability to one.  We can use any function of $y$ for the other $f_i$ according to the appropriate constraints for each problem. For example, if $f_i(y) = y$, then we constrain the final distribution to have mean $\ang{y}$.

To find the maximum entropy distribution, we maximize
	$$\L = S - \psi P - \sum_{i=1}^n \l_iC_i$$
by differentiating with respect to $p_y$, setting to zero, and solving.  This calculation yields
\begin{equation}\label{genMaxEnt}
	p_y = km_ye^{-\sum \l_if_i}, 
\end{equation}
where we choose $k$ so that the total probability is one: $\sum_y p_y=1$ or in the continuous limit $\int_y p_y\dy=1$.  For the additional $n$ constraints, we choose each $\l_i$ so that $\sum_y f_i(y)p_y=\ang{f_i(y)}$ for all $i=1,2,\ldots,n$, using integration rather than summation in the continuous limit.

To solve a particular problem, we must choose a proper measure $m_y$.  In the binomial and Poisson cases, we found $m_y$ by first considering the underlying uniform scale, in which each outcome is assumed to be equally likely in the absence of additional information.  We then calculated the relative weighting of each measure on the $y$ scale, in which, for each $y$, variable numbers of outcomes on the underlying uniform scale map to that value of $y$.  That approach for calculating the transformation from underlying and unobserved uniform measures to observable non-uniform measures commonly solves sampling problems based on combinatorics.  

In continuous cases, we must choose $m_y$ to account for the nature of information provided by measurement.  Notationally, $m_y$ means a function $m$ of the value $y$, alternatively written as $m(y)$.  However, I use subscript notation to obtain an equivalent and more compact expression.  

We find $m_y$ by asking in what ways would changes in measurement leave unchanged the information that we obtain.  That invariance under transformation of the measured scale expresses the lack of information obtained from measurement. Our analysis must always account for the presence or absence of particular information.

Suppose that we cannot directly observe $y$; rather, we can observe only a transformed variable $x=g(y)$.  Under what conditions do we obtain the same information from direct measurement of $y$ or indirect measurement of the transformed value $x$?  Put another way, how can we choose $m$ so that the information we obtain is invariant to certain transformations of our measurements?   

Consider small increments in the direct and transformed scales, $\dy$ and $\dx$.  If we choose $m$ so that $m_y\dy$ is proportional to $m_x\dx$, then our measure $m$ contains proportionally the same increment on both the $y$ and $x$ scales. With a measure $m$ that satisfies this proportionality relation, we will obtain the same maximum entropy probability distribution for both the direct and indirect measurement scales. Thus, we must find an $m$ that satisfies 
\begin{equation}\label{measureEquiv}
	m_y\dy=\k m_x\dx 
\end{equation}
for any arbitrary constant $\k$. The following sections give particular examples.

\subsection*{The exponential distribution}

Suppose we are measuring a positive value such as time or distance.  In this section, I analyze the case in which the average value of observations summarizes all of the information available in the data about the distribution.  Put another way, for a positive variable, suppose the only known constraint in \Eq{genMaxEnt} is the mean: $f_1=y$.  Then from \Eq{genMaxEnt}, 
	$$p_y = km_ye^{-\l_1y}$$
for $y>0$.

We choose $m_y$ to account for the information that we have about the nature of measurement.  In this case, $y$ measures a relative linear displacement in the following sense.  Let $y$ measure the passage of time or a spatial displacement.  Add a constant, $c$, to all of our measurements to make a new measurement scale such that $x=g(y)=y+c$.  Consider the displacement between two points on each scale: $x_2-x_1 = y_2+c-y_1-c = y_2-y_1$.  Thus, relative linear displacement is invariant to arbitrary linear displacement, $c$.  Now consider a uniform stretching $(a>1)$ or shrinking $(a<1)$ of our measurement scale, such that $x=g(y)=ay+c$.  Displacement between two points on each scale is $x_2-x_1 = ay_2+c-ay_1-c = a(y_2-y_1)$. In this case, relative linear displacement changes only by a constant factor between the two scales.

Applying the rules of calculus to $ay+c=x$, increments on the two scales are related by $a\dy=\dx$.  Thus, we can choose $m_y=m_x=1$ and $\k=1/a$ to satisfy \Eq{measureEquiv}.

Using $m_y=1$, we next choose $k$ so that $\int p_y\dy=1$, which yields $k=\l_1$.  To find $\l_1$, we solve $\int y\l_1e^{-\l_1y}\dy=\ang{y}$. Setting $\ang{y}=1/\m$, we obtain $\l_1=\m$.  These substitutions for $k$, $m_y$, and $\l_1$ define the exponential probability distribution
	$$p_y =\m e^{-\m y},$$
where $1/\m$ is the expected value of $y$, which can be interpreted as the average linear displacement.  Thus, if the entire information in a sample about the probability distribution of relative linear displacement is contained in the average displacement, then the most probable or maximum entropy distribution is exponential.  The exponential pattern is widely observed in nature.

\subsection*{The power law distribution}

In the exponential example, we can think of the system as measuring deviations from a fixed point.  In that case, the information in our measures with respect to the underlying probability distribution does not change if we move the whole system---both the fixed point and the measured points---to a new location, or if we uniformly stretch or shrink the measurement scale by a constant factor.  For example, we may measure the passage of time from now until something happens.  In this case, ``now'' can take on any value to set the origin or location for a particular measure. 

By contrast, suppose the distance that we measure between points stretches or shrinks in proportion to the original distance, yet such stretching or shrinking does not change the information that we obtain about the underlying probability distribution.  The invariance of the probability distribution to nonuniform stretching or shrinking of distances between measurements provides information that constrains the shape of the distribution.  We can express this constraint by two measurements of distance or time, $y_1$ and $y_2$, with ratio $y_2/y_1$.  Invariance of this ratio is captured by the transformation $x=y^a$. This transformation yields ratios on the two scales as $x_2/x_1=(y_2/y_1)^a$.  Taking the logarithms of both sides gives $\log(x_2)-\log(x_1)=a[\log(y_2)-\log(y_1)]$; thus, displacements on a logarithmic scale remain the same apart from a constant scaling factor $a$.  

This calculation shows that preserving ratios means preserving logarithmic displacements, apart from uniform changes in scale.  Thus, we fully capture the invariance of ratios by measuring the average logarithmic displacement in our sample.  Given the average of the logarithmic measures, we can apply the same analysis as the previous section, but on a logarithmic scale.  The average logarithmic value is the log of the geometric mean, $\ang{\log(y)}=\log(G)$, where $G$ is the geometric mean. Thus, the only information available to us is the geometric mean of the observations or, equivalently, the average logarithm of the observations, $\ang{\log(y)}$.

We get $m_y$ by examining the increments on the two scales for the transformation $x=y^a$, yielding $\dx=ay^{a-1}\dy$.  If we define the function $m_y=m(y)=1/y$, and apply that function to $x$ and $y^a$, we get from \Eq{measureEquiv}
	$$\ovr{ay^{a-1}}{y^a}\dy=a\ovr{\dy}{y}=\k\ovr{\dx}{x},$$
which means $\dd\log(y)\propto\dd\log(x)$, where $\propto$ means ``proportional to.'' (Note that, in general, $\dd\log(z)=\dd z/z$.) This proportionality confirms the invariance on the logarithmic scale and supports use of the geometric mean for describing information about ratios of measurements.  Because changes in logarithms measure percentage changes in measurements, we can think of the information in terms of how perturbations cause percentage changes in observations.  

From the general solution in \Eq{genMaxEnt}, we use for this problem $m_y=1/y$ and $f_1=\log(y)$, yielding
	$$p_y = (k/y)e^{-\l_1\log(y)}=(k/y)y^{-\l_1}=ky^{-(1+\l_1)}.$$
Power law distributions typically hold above some lower bound, $L\ge0$. I derive the distribution of $1\le L<y<\infty$ as an example.  From the constraint that the total probability is one
	$$\int_L^\infty ky^{-(1+\l_1)}\dy = kL^{-\l_1}/\l_1=1,$$
yielding $k=\l_1L^{\l_1}$.  Next we solve for $\l_1$ by using the constraint on $\ang{\log(y)}$ to write
\begin{align}\notag
			\ang{\log(y)}&=\int_L^\infty \log(y)p_y\dy\notag\\
						&=\int_L^\infty \log(y)\l_1L^{\l_1}y^{-(1+\l_1)}\dy\notag\\
						&=\log(L) + 1/\l_1.\notag
\end{align}
Using $\d=\l_1$, we obtain $\d=1/\ang{\log(y/L)},$ yielding
\begin{equation}\label{powerLaw}
	p_y = \d L^\d y^{-(1+\d)}. 
\end{equation}
If we choose $L=1$, then
	$$p_y = \d y^{-(1+\d)},$$
where $1/\d$ is the geometric mean of $y$ in excess of the lower bound $L$.  Note that the total probability in the upper tail is $(L/y)^\d$.  Typically, one only refers to power law or ``fat tails'' for $\d<2$.

\subsection*{Power laws, entropy, and constraint}

There is a vast literature on power laws.  In that literature, almost all derivations begin with a particular neutral generative model, such as Simon's \citeyear{Simon55functions} preferential attachment model for the frequency of words in languages (see above).  By contrast, I showed that a power law arises simply from an assumption about the measurement scale and from information about the geometric mean.  This view of the power law shows the direct analogy with the exponential distribution: setting the geometric mean attracts aggregates toward a power law distribution; setting the arithmetic mean attracts aggregates toward an exponential distribution.  This sort of informational derivation of the power law occurs in the literature \cite<e.g.,>{Kapur89Engineering,Kleiber03Sciences}, but appears rarely and is almost always ignored in favor of specialized generative models.  

Recently, much work in theoretical physics attempts to find maximum entropy derivations of power laws \cite<e.g.,>{Abe00theorem} from a modified approach called Tsallis entropy \cite{Tsallis88statistics,Tsallis99connections}.  The Tsallis approach uses a more complex definition of entropy but typically applies a narrower concept of constraint than I use in this paper.  Those who follow the Tsallis approach apparently do not accept a constraint on the geometric mean as a natural physical constraint, and seek to modify the definition of entropy so that they can retain the arithmetic mean as the fundamental constraint of location.  

Perhaps in certain physical applications it makes sense to retain a limited view of physical constraints.  But from the broader perspective of pattern, beyond certain physical applications, I consider the geometric mean as a natural informational constraint that arises from measurement or assumption. By this view, the simple derivation of the power law given here provides the most general outlook on the role of information in setting patterns of nature.  

\subsection*{The gamma distribution}

If the average displacement from an arbitrary point captures all of the information in a sample about  the probability distribution, then observations follow the exponential distribution.  If the average logarithmic displacement captures all of the information in a sample, then observations follow the power law distribution. Displacements are nonnegative values measured from a reference point.  

In this section, I show that if the average displacement and the average logarithmic displacement together contain all the information in a sample about the underlying probability distribution, then the observations follow the gamma distribution.

No transformation preserves the information in both direct and logarithmic measures apart from uniform scaling, $x=ay$.  Thus, $m_y$ is a constant and drops out of the analysis in the general solution given in \Eq{genMaxEnt}.  From the general solution, we use the constraint on the mean, $f_1=y$, and the constraint on the mean of the logarithmic values, $f_2=\log(y)$, yielding
	$$p_y = ke^{-\l_1 y - \l_2\log(y)}=ky^{-\l_2}e^{-\l_1 y}.$$
We solve for the three unknowns $k$, $\l_1$, and $\l_2$ from the constraints on the total probability, the mean, and the mean logarithmic value (geometric mean).  For convenience, make the substitutions $\m=\l_1$ and $r=1-\l_2$.  Using each constraint in turn and solving for each of the unknowns yields the gamma distribution
	$$p_y = \ovr{\m^r}{\Gamma(r)}y^{r-1}e^{-\m y},$$
where $\Gamma$ is the gamma function, the average value is $\ang{y}=r/\m$, and the average logarithmic value is $\ang{\log(y)}=-\log(\m)+\Gamma'(r)/\Gamma(r)$, where the prime denotes differentiation with respect to $r$. Note that the gamma distribution is essentially a product of a power law, $y^{r-1}$, and an exponential, $e^{-\m y}$, representing the combination of the independent constraints on the geometric and arithmetic means.

The fact that both linear and logarithmic measures provide information suggests that measurements must be made in relation to an absolute fixed point.  The need for full information of location may explain why the gamma distribution often arises in waiting time problems, in which the initial starting time denotes a fixed birth date that sets the absolute location of measure.

\subsection*{The Gaussian distribution}

Suppose one knows the mean, $\m$, and the variance, $\s^2$, of a population from which one makes a set of measurements.  Then one can express a measurement, $y$, as the deviation $x=(y-\m)/\s$, where $\s$ is the standard deviation.  One can think of $1/\s^2$ as the amount of information one obtains from an observation about the location of the mean, because the smaller $\s^2$, the closer observed values $y$ will be to $\m$.  

If all one knows is $1/\s^2$, the amount of information about location per observation, then the probability distribution that expresses that state of knowledge is the Gaussian (or normal) distribution.  If one has no information about location, $\m$, then the most probable distribution centers at zero, expressing the magnitude of fluctuations.  If one knows the location, $\m$, then the most probable distribution is also a Gaussian with the same shape and distribution of fluctuations, but centered at $\m$.

The widespread use of Gaussian distributions arises for two reasons.  First, many measurements concern fluctuations about a central location caused by perturbing factors or by errors in measurement.  Second, in formulating a theoretical analysis of measurement and information, an assumption of Gaussian fluctuations is the best choice when one has information only about the precision or error in observations with regard to the average value of the population under observation \cite{Jaynes03Science}.

The derivation of the Gaussian follows our usual procedure.  We assume that the mean, $\ang{y}=\m$, and the variance, $\ang{(y-\m)^2}=\s^2$, capture all of the information in observations about the probability distribution.  Because the mean enters only through the deviations $y-\m$, we need only one constraint from \Eq{genMaxEnt} expressed as $f_1=(y-\m)^2$.  With regard to $m_y$, the expression $x=(y-\m)/\s$ captures the invariance under which we lose no information about the distribution.  Thus, $\dx=\dy/\s$, leads to a constant value for $m_y$ that drops out of the analysis.  From \Eq{genMaxEnt}, 
	$$p_y = ke^{-\l_1(y-\m)^2}.$$
We find $k$ and $\l_1$ by solving the two constraints $\int p_y\dy=1$ and $\int (y-\m)^2p_y\dy=\s^2$.  Solving gives $k^{-1}=\s\sqrt{2\pi}$ and $\l_1^{-1}=2\s^2$, yielding the Gaussian distribution
\begin{equation}\label{maxEntGauss}
	p_y = \ovr{1}{\s\sqrt{2\pi}}\,e^{-(y-\m)^2/2\s^2}, 
\end{equation}
or expressed more simply in terms of the normalized deviations $x=(y-\m)/\s$ as
	$$p_x = \ovr{1}{\sqrt{2\pi}}\,e^{-x^2/2}.$$

\section*{Limiting distributions}

Most observable patterns of nature arise from aggregation of numerous small scale processes.  I have emphasized that aggregation tends to smooth fluctuations, so that the remaining pattern converges to maximum entropy subject to the constraints of the information or signal that remains.  We might say that, as the number of entities contributing to the aggregate increases, we converge in the limit to those maximum entropy distributions that define the common patterns of nature.

In this section, I look more closely at the process of aggregation.  Why do fluctuations tend to cancel in the aggregate?  Why is aggregation so often written as a summation of observations?  For example, the central limit theorem is about the way in which a sum of observations converges to a Gaussian distribution as we increase the number of observations added to the sum.  Similarly, I discussed the binomial distribution as arising from the sum of the number of successes in a series of independent trials, and the Poisson distribution as arising from the number of counts of some event summed over a large number of small temporal or spatial intervals.  

It turns out that summation of random variables is really a very general process that smooths the fluctuations in observations. Such smoothing very often acts as a filter to remove the random noise that lacks a signal and to enhance the true signal or information contained in the aggregate set of observations.  Put another way, summation of random processes is much more than our usual intuitive concept of simple addition.

I mentioned that we already have encountered the binomial and Poisson distributions as arising from summation of many independent observations.  Before I turn to general aspects of summation, I first describe the central limit theorem in which sums of random variables often converge to a Gaussian distribution.

\section*{The central limit theorem and the Gaussian distribution}

\vskip-18pt
\begin{quotation}
\textit{A Gaussian probability distribution has higher entropy than any other with the same variance; therefore any operation on a probability distribution which discards information, but conserves variance, leads us inexorably closer to a Gaussian. The central limit theorem $\ldots$ is the best known example of this, in which the operation performed is convolution [summation of random processes] \cite[p.~221]{Jaynes03Science}. }
\end{quotation}

A combination of random fluctuations converges to the Gaussian if no fluctuation tends to dominate.  The lack of dominance by any particular fluctuation is what Jaynes means by ``conserves variance''; no fluctuation is too large as long as the squared deviation (variance) for that perturbation is not, on average, infinitely large relative to the other fluctuations.  

One encounters in the literature many special cases of the central limit theorem.  The essence of each special case comes down to information.  Suppose some process of aggregation leads to a probability distribution that can be observed.  If all of the information in the observations about the probability distribution is summarized completely by the variance, then the distribution is Gaussian.  We ignore the mean, because the mean pins the distribution to a particular location, but does not otherwise change the shape of the distribution.  

Similarly, suppose the variance is the only constraint we can assume about an unobserved distribution---equivalently, suppose we know only the precision of observations about the location of the mean, because the variance defines precision.  If we can set only the precision of observations, then we should assume the observations come from a Gaussian distribution.  

We do not know all of the particular generative processes that converge to the Gaussian.  Each particular statement of the central limit theorem provides one specification of the domain of attraction---a subset of the generative models that do in the limit take on the Gaussian shape.  I briefly mention three forms of the central limit theorem to a give a sense of the variety of expressions.

First, for any random variable with finite variance, the sum of independent and identical random variables converges to a Gaussian as the number of observations in the sum increases.  This statement is the most common in the literature. It is also the least general, because it requires that each observation in the sum come from the same identical distribution, and that each observation be independent of the other observations.

Second, the Lindeberg condition does not require that each observation come from the same identical distribution, but it does require that each observation be independent of the others and that the variance be finite for each random variable contributing to the sum \cite{Feller71Applications}.  In practice, for a sequence of $n$ measurements with sum $Z_n=\sum X_i/\sqrt{n}$ for $i=1,\ldots,n$, and if $\s^2_i$ is the variance of the $i$th variable so that $V_n=\sum \s^2_i/n$ is the average variance, then $Z_n$ approaches a Gaussian as long as no single variance $\s^2_i$ dominates the average variance $V_n$.  

Third, the martingale central limit theorem defines a generative process that converges to a Gaussian in which the random variables in a sequence are neither identical nor independent \cite{Hall80Application}.  Suppose we have a sequence of observations, $X_t$, at successive times $t=1,\ldots,T$. If the expected value of each observation equals the value observed in the prior time period, and the variance in each time period, $\s^2_t$, remains finite, then the sequence $X_t$ is a martingale that converges in distribution to a Gaussian as time increases.  Note that the distribution of each $X_t$ depends on $X_{t-1}$; the distribution of $X_{t-1}$ depends on $X_{t-2}$; and so on. Therefore each observation depends on all prior observations.   

Extension of the central limit theorem remains a very active field of study \cite{Johnson04Theorem}.  A deeper understanding of how aggregation determines the patterns of nature justifies that effort.

In the end, information remains the key.  When all information vanishes except the variance, pattern converges to the Gaussian distribution.  Information vanishes by repeated perturbation. Variance and precision are equivalent for a Gaussian distribution: the information (precision) contained in an observation about the average value is the reciprocal of the variance, $1/\s^2$. So we may say that the Gaussian distribution is the purest expression of information or error in measurement \cite{Stigler861900}.

As the variance goes to infinity, the information per observation about the location of the average value, $1/\s^2$, goes to zero.  It may seem strange that an observation could provide no information about the mean value.  But some of the deepest and most interesting aspects of pattern in nature can be understood by considering the Gaussian distribution to be a special case of a wider class of limiting distributions with potentially infinite variance.  

When the variance is finite, the Gaussian pattern follows, and observations provide information about the mean.  As the variance becomes infinite because of occasional large fluctuations, one loses all information about the mean, and patterns follow a variety of power law type distributions.  Thus, Gaussian and power law patterns are part of a single wider class of limiting distributions, the \levy\ stable distributions.  Before I turn to the \levy\ stable distributions, I must develop the concept of aggregation more explicitly.

\section*{Aggregation: summation and its meanings}

Our understanding of aggregation and the common patterns in nature arises mainly from concepts such as the central limit theorem and its relatives. Those theorems tell us what happens when we sum up random processes.  

Why should addition be the fundamental concept of aggregation?  Think of the complexity in how processes combine to form the input-output relations of a control network, or the complexity in how numerous processes influence the distribution of species across a natural landscape.

Three reasons support the use of summation as a common form of aggregation.  First, multiplication and division can be turned into addition or subtraction by taking logarithms.  For example, the multiplication of numerous processes often smooths into a Gaussian distribution on the logarithmic scale, leading to the log-normal distribution.  

Second, multiplication of small perturbations is roughly equivalent to addition.  For example, suppose we multiply two processes each perturbed by a small amount, $\e$ and $\d$, respectively, so that the product of the perturbed processes is $(1+\e)(1+\d) = 1+\e+\d+\e\d\approx 1+\e+\d$.  Because $\e$ and $\d$ are small relative to one, their product is very small and can be ignored.  Thus, the total perturbations of the multiplicative process are simply the sum of the perturbations.  In general, aggregations of small perturbations combine through summation.

Third, summation of random processes is rather different from a simple intuitive notion of adding two numbers.  Instead, adding a stochastic variable to some input acts like a filter that typically smooths the output, causing loss of information by taking each input value and smearing that value over a range of outputs.  Therefore, summation of random processes is a general expression of perturbation and loss of information. With an increasing number of processes, the aggregate increases in entropy toward the maximum, stable value of disorder defined by the sampling structure and the information preserved through the multiple rounds of perturbations.

The following two subsections give some details about adding random processes.  These details are slightly more technical than most of the paper; some readers may prefer to skip ahead.  However, these details ultimately reveal the essence of pattern in natural history, because pattern in natural history arises from aggregation.

\subsection*{Convolution: the addition of random processes}

Suppose we make two independent observations from two random processes, $X_1$ and $X_2$.  What is the probability distribution function (pdf) of the sum, $X=X_1+X_2$?

\begin{figure}[htp]
\centering
\includegraphics[totalheight=0.5\textheight,viewport=0 0 2200 2600,clip]{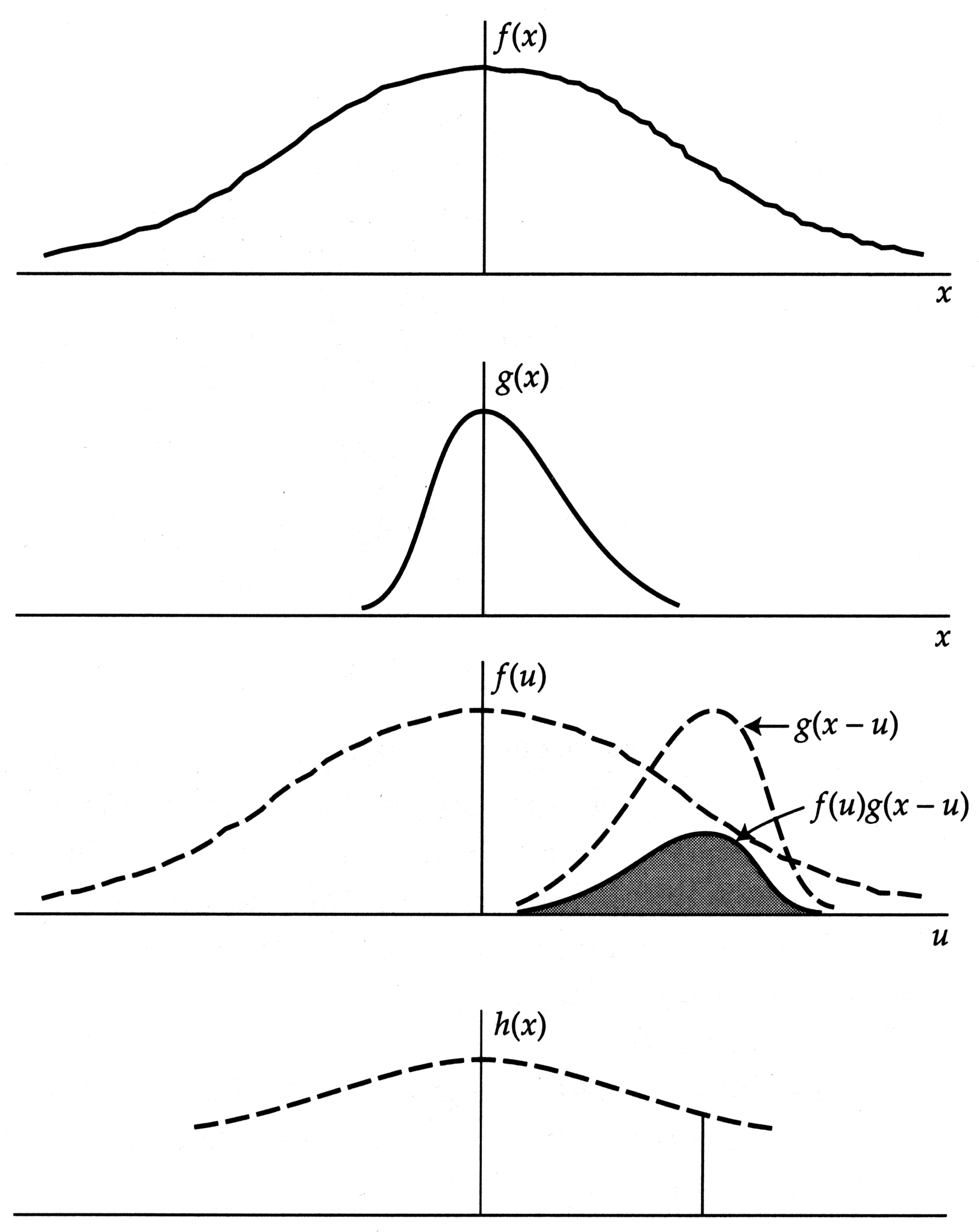}
\caption[] {Summing two independent random variables smooths the distribution of the sum. The plots illustrate the process of convolution given by \Eq{convDef}. The top two plots show the separate distributions of $f$ for $X_1$ and $g$ for $X_2$.  Note that the initial distribution of $X_1$ given by $f$ is noisy; one can think of adding $X_2$ to $X_1$ as applying the filter $g$ to $f$ to smooth the noise out of $f$.  The third plot shows how the smoothing works at an individual point marked by the vertical bar in the lower plot.  At that point, $u$, in the third plot, the probability of observing $u$ from the initial distribution is proportional to $f(u)$. To obtain the sum, $x$, the value from the second distribution must be $x-u$, which occurs with probability proportional to $g(x-u)$.  For each fixed $x$ value, one obtains the total probability $h(x)$ in proportion to the sum (integral) over $u$ of all the different $f(u)g(x-u)$ combinations, given by the shaded area under the curve. From \citeA[figure 3.1]{Bracewell00Applications}.\label{fig:bwConv1}}
\end{figure}

\begin{figure}[htp]
\centering
\includegraphics[totalheight=0.5\textheight,viewport=0 0 300 350,clip]{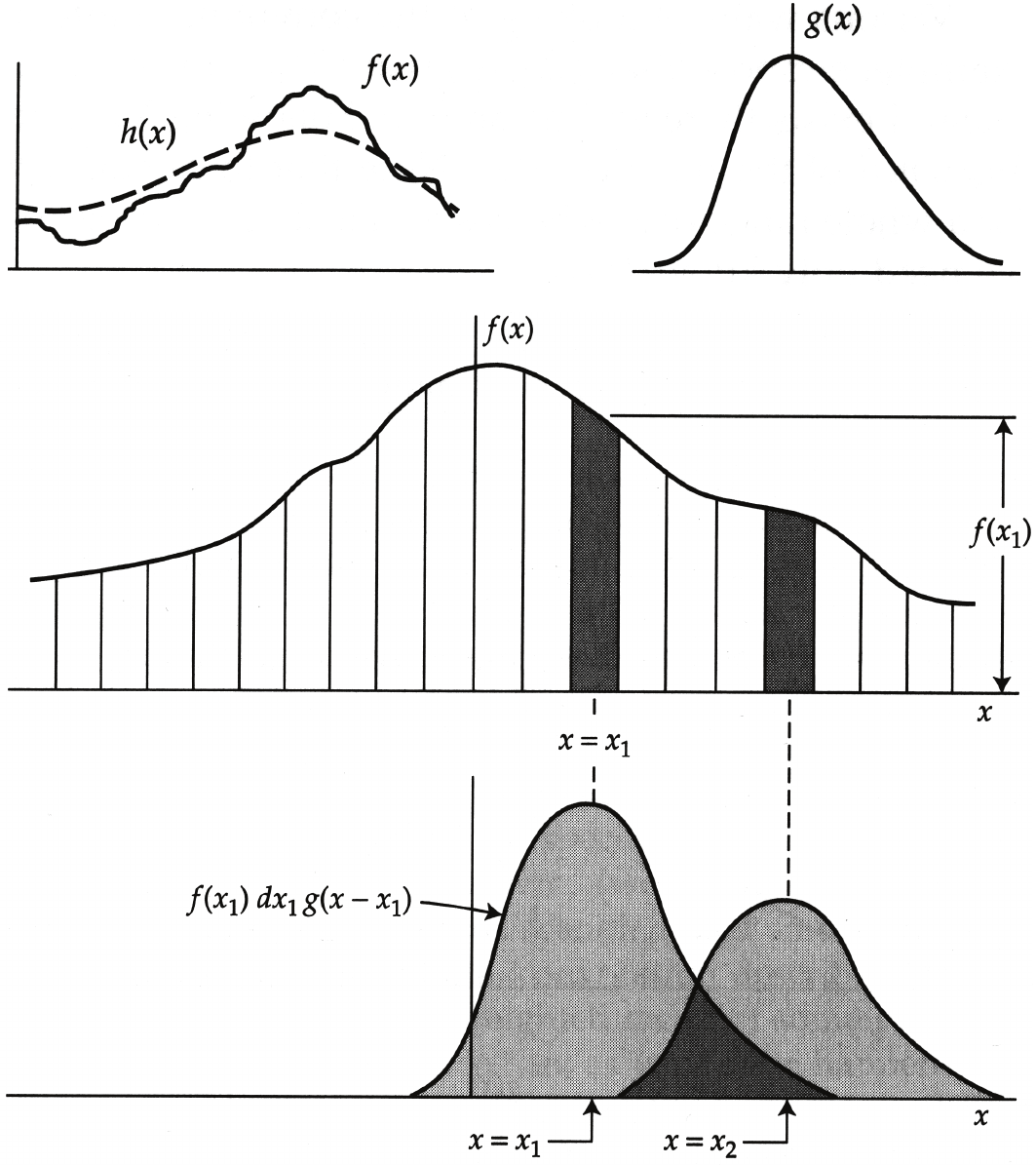}
\caption[] {Another example of how summing random variables (convolution) smooths a distribution.  The top plots show the initial noisy distribution $f$ and a second, smoother distribution, $g$. The distribution of the sum, $h=f*g$, smooths the initial distribution of $f$. The middle plot shows a piece of $f$ broken into intervals, highlighting two intervals $x=x_1$ and $x=x_2$.  The lower panel shows how convolution of $f$ and $g$ gives the probability, $h(x)$, that the sum takes on a particular value, $x$. For example, the value $h(x_1)$ is the shaded area under the left curve, which is the sum (integral) of $f(u)g(x-u)$ over all values of $u$ and then evaluated at $x=x_1$. The area under the right curve is $h(x_2)$ obtained by the same calculation evaluated at $x=x_2$. From \citeA[figures 3.2 and 3.3]{Bracewell00Applications}.\label{fig:bwConv2}}
\end{figure}

Let $X_1$ have pdf $f(x)$ and $X_2$ have pdf $g(x)$.  Then the pdf of the sum, $X=X_1+X_2$, is 
\begin{equation}\label{convDef}
	h(x) = \int f(u)g(x-u)\dd u. 
\end{equation}
Read this as follows: for each possible value, $u$, that $X_1$ can take on, the probability of observing that value is proportional to $f(u)$.  To obtain the sum, $X_1+X_2=x$, given that $X_1=u$, it must be that $X_2=x-u$, which occurs with probability $g(x-u)$.  Because $X_1$ and $X_2$ are independent, the probability of $X_1=u$ and $X_2=x-u$ is $f(u)g(x-u)$.  We then add up (integrate over) all combinations of observations that sum to $x$, and we get the probability that the sum takes on the value $X_1+X_2=x$.  Figures \ref{fig:bwConv1} and \ref{fig:bwConv2} illustrate how the operation in \Eq{convDef} smooths the probability distribution for the sum of two random variables.

The operation in \Eq{convDef} is called convolution: we get the pdf of the sum by performing convolution of the two distributions for the independent processes that we are adding.  The convolution operation is so common that it has its own standard notation: the distribution, $h$, of the sum of two independent random variables with distributions $f$ and $g$, is the convolution of $f$ and $g$, which we write as
\begin{equation}\label{convNotate}
	h=f*g. 
\end{equation}
This notation is just a shorthand for \Eq{convDef}.  

\subsection*{The Fourier transform: the key to aggregation and pattern}

The previous section emphasized that aggregation often sums random fluctuations.  If we sum two independent random processes, $Y_2=X_1+X_2$, each drawn from the same distribution, $f(x)$, then the distribution of the sum is the convolution of $f$ with itself: $g(x)=f*f=f^{*2}$.  Similarly, if we summed $n$ independent observations from the same distribution, 
\begin{equation}\label{sumY}
	Y_n=\sum_{i=1}^nX_i, 
\end{equation}
then $g(x)$, the distribution of $Y_n$, is the $n$-fold convolution $g(x)=f^{*n}$. Thus, it is very easy, in principle, to calculate the distribution of a sum of independent random fluctuations.  However, convolution given by \Eq{convDef} is tedious and does not lead to easy analysis.  

Fourier transformation provides a useful way to get around the difficulty of multiple convolutions.  Fourier transformation partitions any function into a combination of terms, each term describing the intensity of fluctuation at a particular frequency.   Frequencies are a more natural scale on which to aggregate and study fluctuations, because weak signals at particular frequencies carry little information about the true underlying pattern and naturally die away upon aggregation.  

To show how Fourier transformation extinguishes weak signals upon aggregation of random fluctuations, I start with the relation between Fourier transformation and convolution.  The Fourier transform takes some function, $f(x)$, and changes it into another function, $F(s)$, that contains exactly the same information but expressed in a different way.  In symbols, the Fourier transform is
	$$\F\{f(x)\}=F(s).$$
The function $F(s)$ contains the same information as $f(x)$, because we can reverse the process by the inverse Fourier transform
	$$\Finv\{F(s)\}=f(x).$$

We typically think of $x$ as being any measurement such as time or distance; the function $f(x)$ may, for example, be the pdf of $x$, which gives the probability of a fluctuation of magnitude $x$.  In the transformed function, $s$ describes the fluctuations with regard to their frequencies of repetition at a certain magnitude, $x$, and $F(s)$ is the intensity of fluctuations of frequency $s$.  We can express fluctuations by sine and cosine curves, so that $F$ describes the weighting or intensity of the combination of sine and cosine curves at frequency $s$.  Thus, the Fourier transform takes a function $f$ and breaks it into the sum of component frequency fluctuations with particular weighting $F$ at each frequency $s$.  I give the technical expression of the Fourier transform at the end of this section.

With regard to aggregation and convolution, we can express a convolution of probability distributions as the product of their Fourier transforms.  Thus, we can replace the complex convolution operation with multiplication.  After we have finished multiplying and analyzing the transformed distributions, we can transform back to get a description of the aggregated distribution on the original scale.  In particular, for two independent distributions $f(x)$ and $g(x)$, the Fourier transform of their convolution is
	$$\F\{(f*g)(x)\}=F(s)G(s).$$
When we add $n$ independent observations from the same distribution, we must perform the $n$-fold convolution, which can also be done by multiplying the transformed function $n$ times
	$$\F\{f^{*n}\}=[F(s)]^n.$$
Note that a fluctuation at frequency $\om$ with weak intensity $F(\om)$ will get washed out compared with a fluctuation at frequency $\om'$ with strong intensity $F(\om')$, because 
	$$\ovr{[F(\om)]^n}{[F(\om')]^n}\to0$$
with an increase in the number of fluctuations, $n$, contributing to the aggregate. Thus the Fourier frequency domain makes clear how aggregation intensifies strong signals and extinguishes weak signals.  

\subsection*{The central limit theorem and the Gaussian distribution}

\begin{figure}[t]
\centering
\includegraphics[totalheight=0.5\textheight,viewport=0 0 240 175,clip]{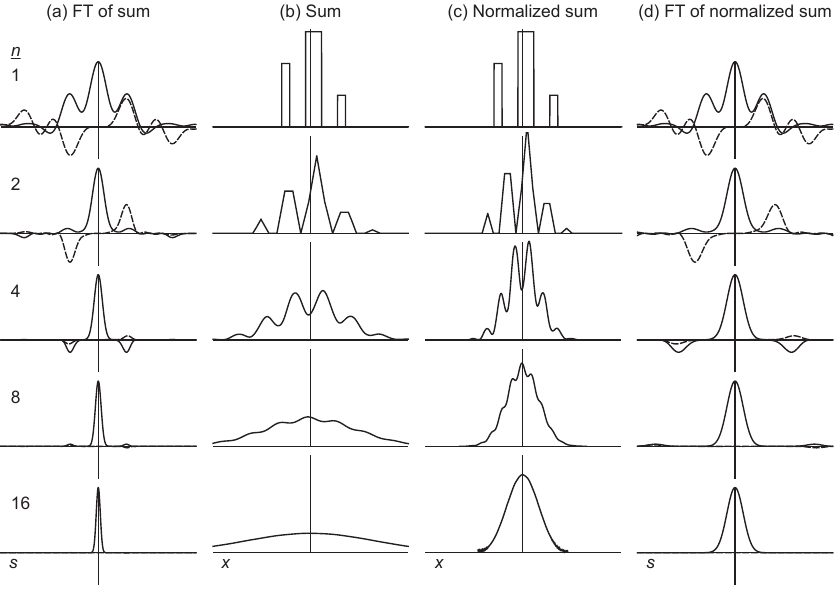}
\caption[] {The powerful smoothing caused by aggregation.  In the Fourier transform (FT) columns, the solid curve is the cosine component of the transform, and dashed curve is the sine component of the transform from \Eq{fourierDef}.\label{fig:CLT}}
\end{figure}

Figure \ref{fig:CLT} illustrates how aggregation cleans up signals in the the Fourier domain.  The top panel of column (b) in the figure shows the base distribution $f(x)$ for the random variable $X$.  I chose an idiosyncratic distribution to demonstrate the powerful smoothing effect upon aggregation:
\begin{align}
	f(x) =  
	\begin{cases}			
							0.682\mskip30mu &\hbox{if $-0.326 < x < \phantom{-}0.652$}\notag\\
							0.454\mskip30mu	&\hbox{if $-1.793 < x < -1.304$}\\
							0.227\mskip30mu &\hbox{if $\phantom{-}1.630 < x < \phantom{-}2.119$.}\notag\\
	\end{cases}\label{idioF}\\
\end{align}
The distribution $f$ has a mean $\m=0$ and a variance $\s^2=1$.  

Column (b) of the figure shows the distribution $g(x)$ of the sum
	$$Y_n=\sum_{i=1}^n X_i,$$
where the $X_i$ are independent and distributed according to $f(x)$ given in \Eq{idioF}.  The rows show increasing values of $n$.  For each row in column (b), the distribution $g$ is the $n$-fold convolution of $f$, that is, $g(x)=f^{*n}(x)$.  Convolution smooths the distribution and spreads it out; the variance of $Y_n$ is $n\s^2$, where in this case the base variance is $\s^2=1$.

Column (a) shows the Fourier transform of $g(x)$, which is $G(s)=[F(s)]^n$, where $F$ is the Fourier transform of $f$. The peak value is at $F(0)=1$, so for all other values of $s$, $F(s) < 1$, and the value $[F(s)]^n$ declines as $n$ increases down the rows.  As $n$ increases, the Fourier spectrum narrows toward a peak at zero, while the distribution of the sum in column (b) continues to spread more widely.  This corresponding narrowing in the Fourier domain and widening in the direct domain go together, because a spread in the direct domain corresponds to greater intensity of wide, low frequency signals contained in the spreading sum.

The narrowing in column (a) and spreading in column (b) obscure the regularization of shape that occurs with aggregation, because with an increasing number of terms in the sum, the total value tends to fluctuation more widely.  We can normalize the sum to see clearly how the shape of the aggregated distribution converges to the Gaussian by the central limit theorem.  Write the sum as
	$$Z_n = \ovr{1}{\sqrt{n}}\sum_{i=1}^n X_i=Y_n\big/\sqrt{n},$$
and define the distribution of the normalized sum as $h(x)$.  With this normalization, the variance of $Z_n$ is $\s^2=1$ independently of $n$, and the distribution $h(x) = \sqrt{n}g(x/\sqrt{n})$.  This transformation describes the change from the plot of $g$ in column (b) to $h$ in column (c).  In particular, as $n$ increases, $h$ converges to the Gaussian form with zero mean
	$$h(x) = \ovr{1}{\sqrt{2\pi\s^2}}\;e^{-x^2/2\s^2}.$$

Column (d) is the Fourier transform, $H(s)$, for the distribution of the standardized sum, $h(x)$.  The Fourier transform of the unstandardized sum in column (a) is $G(s)$, and $H(s) = G(s/\sqrt{n})$.  Interestingly, as $n$ increases, $H(s)$ converges to a Gaussian shape
\begin{equation}\label{gaussFourier}
	H(s) = e^{-\g^2 s^2}, 
\end{equation}
in which $\g^2=\s^2/2$. The Gaussian is the only distribution which has a Fourier transform with the same shape.

\subsection*{Maximum entropy in the Fourier domain}

The direct and Fourier domains contain the same information.  Thus, in deriving most likely probability distributions subject to the constraints of the information that we have about a particular problem, we may work equivalently in the direct or Fourier domains.  In the direct domain, we have applied the method of maximum entropy throughout the earlier sections of this paper.  Here, I show the equivalent approach to maximizing entropy in the Fourier domain.

To obtain the maximum entropy criterion in the Fourier domain \cite{Johnson84S}, we need an appropriate measure of entropy by analogy with \Eq{KLentropy}.  To get a probability distribution from a Fourier domain function, $H(s)$, we normalize so that the total area under the Fourier curve is one
	$$H'(s) = H(s)\bigg/\int_{-\infty}^\infty H(s)\ds.$$
For simplicity, I assume that the direct corresponding distribution, $h(x)$, is centered at zero and is symmetric, so that $H$ is symmetric and does not have an imaginary component.  If so, then the standardized form of $H'$ given here is an appropriate, symmetric pdf.  With a pdf defined over the frequency domain, $s$, we can apply all of the standard tools of maximum entropy developed earlier.  The corresponding equation for entropy by reexpressing \Eq{KLentropy} is
\begin{equation}\label{FourierEntropy}
	S=-\int H'(s) \log\left[\ovr{H'(s)}{M(s)}\right]\ds, 
\end{equation}
where $M(s)$ describes prior information about the relative intensity (probability distribution) of frequencies, $s$.  Prior information may arise from the sampling structure of the problem or from other prior information about the relative intensity of frequencies.  

With a definition of entropy that matches the standard form used earlier, we can apply the general solution to maximum entropy problems in \Eq{genMaxEnt}, which we write here as
\begin{equation}\label{fourierMaxEnt}
	H'(s) = kM(s)e^{-\sum \l_if_i}, 
\end{equation}
where we choose $k$ so that the total probability is one: $\int H'(s)\ds=1$.  For the additional $n$ constraints, we choose each $\l_i$ so that $\int f_i(s)H'(s)\ds=\ang{f_i(s)}$ for $i=1,2,\ldots,n$.  For example, if we let $f_1(s)=s^2$, then we constrain the second moment (variance) of the spectral distribution of frequencies.  Spectral moments summarize the location and spread of the power concentrated at various frequencies \shortcite{Cramer67Applications,Benoit92method}.  Here, we can assume $M(s)=1$, because we have no prior information about the spectral distribution.  Thus,
	$$H'(s)=ke^{-\l_1 s^2}.$$
We need to choose $k$ so that $\int H'(s)\ds=1$ and choose $\l_1$ so that $\int s^2H'(s)\ds=\ang{s^2}$.  The identical problem arose when using maximum entropy to derive the Gaussian distribution in \Eq{maxEntGauss}.  Here, we have assumed that $s$ is symmetric and centered at zero, so we can take the mean to be zero, or from \Eq{maxEntGauss}, $\m=0$.  Using that solution, we have
	$$H'(s) = ke^{-s^2/2\ang{s^2}},$$
where $\ang{s^2}$ is the spectral variance.  It turns out that, if we denote the variance of the direct distribution $h(x)$ as $\s^2$, then $\ang{s^2}=1/\s^2$; that is, the spectral variance is the inverse of the direct variance.   Here, let us use $\g^2 = \s^2/2=1/2\ang{s^2}$, so that we keep a separate notation for the spectral distribution.  Then
	$$H'(s) = ke^{-\g^2s^2}.$$
The function $H'$ is a spectral probability distribution that has been normalized so that the probability totals to one.  However, the actual spectrum of frequencies in the Fourier domain, $H(s)$, does not have a total area under its curve of one.  Instead the correct constraint for $H(s)$ is that $H(0)=1$, which constrains the area under the probability distribution on the direct scale, $h(x)$, to total to one.  If we choose $k=1$, we obtain the maximum entropy solution of spectral frequencies in the Fourier domain for the probability distribution $h(x)$, subject to a constraint on the spectral variance $\g^2 = 1/2\ang{s^2}$, as
\begin{equation}\label{spectralMaxEntGauss}
	H(s) = e^{-\g^2s^2}. 
\end{equation}
If we take the inverse Fourier transform of $H(s)$, we obtain a Gaussian distribution $h(x)$.  This method of spectral maximum entropy suggests that we can use information or assumptions about the spectrum of frequencies in the Fourier domain to obtain the most likely probability distributions that describe pattern in the domain of directly observable measurements.  

At first glance, this method of maximum entropy in the frequency domain may seem unnecessarily complicated.  But it turns out that the deepest concepts of aggregation and pattern can only be analyzed in the frequency domain.  The primacy of the frequency domain may occur because of the natural way in which aggregation suppresses minor frequencies as noise and enhances major frequencies as the signals by which information shapes pattern.  I develop these points further after briefly listing some technical details.

\subsection*{Technical details of the Fourier transform}

The Fourier transform is given by 
\begin{equation}\label{fourierDef}
	F(s) = \F\{f(x)\} =\int_{-\infty}^\infty f(x)e^{-ixs}\dx. 
\end{equation}
The frequency interpretation via sine and cosine curves arises from the fact that $e^{is}=\cos(s)+i\sin(s)$.  Thus one can expand the Fourier transform into a series of sine and cosine components expressed in terms of frequency $s$. 

The inverse transformation demonstrates the full preservation of information when changing between $x$ and $s$, given as
	$$f(x) = \Finv\{F(s)\}=\ovr{1}{2\pi}\int_{-\infty}^\infty F(s)e^{ixs}\ds.$$

\section*{The \levy\ stable distributions}

When we sum variables with finite variance, the distribution of the sum converges to a Gaussian.  Summation is one particular generative process that leads to a Gaussian.  Alternatively, we may consider the distributional problem from an information perspective.  If all we know about a distribution is the variance---the precision of information in observations with respect to the mean---then the most likely distribution derived by maximum entropy is Gaussian.  From the more general information perspective, summation is just one particular generative model that leads to a Gaussian.  The generative and information approaches provide distinct and complementary ways in which to understand the common patterns of nature.

In this section, I consider cases in which the variance can be infinite. Generative models of summation converge to a more general form, the \levy\ stable distributions.  The Gaussian is just a special case of the \levy\ stable distributions---the special case of finite variance.   From an information perspective, the \levy\ stable distributions arise as the most likely pattern given knowledge only about the moments of the frequency spectrum in the Fourier domain.  In the previous section, I showed that information about the spectral variance leads, by maximum entropy, to the Gaussian distribution.  In this section, I show that information about other spectral moments, such as the mean of the spectral distribution, leads to other members from the family of \levy\ stable distributions.  The other members, besides the Gaussian, have infinite variance.

When would the variance be infinite?  Perhaps never in reality.  More realistically, the important point is that observations with relatively large values occur often enough that a set of observations provides very little information about the average value.

\subsection*{Large variance and the law of large numbers}

Consider a random variable $X$ with mean $\m$ and variance $\s^2$.  The sum
	$$Z_n=\ovr{1}{n}\sum_{i=1}^nX_i=\bar{X}$$
is the sample mean, $\bar{X}$, for a set of $n$ independent observations of the variable $X$.  If $\s^2$ is finite then, by the central limit theorem, we know that $\bar{X}$ has a Gaussian distribution with mean $\m$ and variance $\s^2/n$.  As $n$ increases, the variance $\s^2/n$ becomes small, and $\bar{X}$ converges to the mean $\m$.  We can think of $\s^2/n$, the  variance of $\bar{X}$, as the spread of the estimate $\bar{X}$ about the the true mean, $\m$.  Thus the inverse of the spread, $n/\s^2$, measures the precision of the estimate.  For finite $\s^2$, as $n$ goes to infinity, the precision $n/\s^2$ also becomes infinite as $\bar{X}$ converges to $\m$.  

If the variance $\s^2$ is very large, then the precision $n/\s^2$ remains small even as $n$ increases.  As long as $n$ does not exceed $\s^2$, precision is low.  Each new observation provides additional precision, or information, about the mean in proportion to $1/\s^2$.  As $\s^2$ becomes very large, the information about the mean per observation approaches zero.

For example, consider the power law distribution for $X$ with pdf $1/x^2$ for $x>1$.  The probability of observing a value of $X$ greater than $k$ is $1/k$.  Thus, any new observation can be large enough to overwhelm all previous observations, no matter how many observations we have already accumulated.  In general, new observations occasionally overwhelm all previous observations whenever the variance is infinite, because the precision added by each observation, $1/\s^2$, is zero.  A sum of random variables, or a random walk, in which any new observation can overwhelm the information about location in previous observations, is called a \levy\ flight.

Infinite variance can be characterized by the total probability in the extreme values, or the tails, of a probability distribution.  For a distribution $f(x)$, if the probability of $|x|$ being greater than $k$ is greater than $1/k^2$ for large $k$, then the variance is infinite.  By considering large values of $k$, we focus on how much probability there is in the tails of the distribution.  One says that when the total probability in the tails is greater than $1/k^2$, the distribution has ``fat tails,'' the variance is infinite, and a sequence follows a \levy\ flight.

Variances can be very large in real applications, but probably not infinite.  Below I discuss truncated \levy\ flights, in which probability distributions have a lot of weight in the tails and high variance.  Before turning to that practical issue, it helps first to gain full insight into the case of infinite variance.  Real cases of truncated \levy\ flights with high variance tend to fall between the extremes of the Gaussian with moderate variance and the \levy\ stable distributions with infinite variance.

\subsection*{Generative models of summation}

Consider the sum
\begin{equation}\label{levySum}
	Z_n = \ovr{1}{n^{1/\a}}\sum_{i=1}^n X_i 
\end{equation}
for independent observations of the random variable $X$ with mean zero.  If the variance of $X$ is finite and equal to $\s^2$, then with $\a=2$ the distribution of $Z_n$ converges to a Gaussian with mean zero and variance $\s^2$ by the central limit theorem.  In the Fourier domain, the distribution of the Gaussian has Fourier transform
\begin{equation}\label{levyFourier}
	H(s) = e^{-\g^\a |s|^\a} 
\end{equation}
with $\a=2$, as given in \Eq{gaussFourier}.  If $\a<2$, then the variance of $X$ is infinite, and the fat tails are given by the total probability $1/|x|^\a$ above large values of $|x|$. The Fourier transform of the distribution of the sum $Z_n$ is given by \Eq{levyFourier}.  The shape of the distribution of $X$ does not matter, as long as the tails follow a power law pattern. 

Distributions with Fourier transforms given by \Eq{levyFourier} are called \levy\ stable distributions.  The full class of \levy\ stable distributions has a more complex Fourier transform with additional parameters for the location and skew of the distribution.  The case given here assumes distributions in the direct domain, $x$, are symmetric with a mean of zero.  We can write the symmetric \levy\ stable distributions in the direct domain only for $\a=2$, which is the Gaussian, and for $\a=1$ which is the Cauchy distribution given by
	$$h(x) = \ovr{\g}{\pi(\g^2+x^2)}.$$
As $(x/\g)^2$ increases above about 10, the Cauchy distribution approximately follows a pdf with  a power law distribution $1/|x|^{1+\a}$, with total probability in the tails $1/|x|^\a$ for $\a=1$.  

In general, the particular forms of the \levy\ stable distributions are known only from the forms of their Fourier transforms.  The fact that the general forms of these distributions have a simple expression in the Fourier domain occurs because the Fourier domain is the natural expression of aggregation by summation. In the direct domain, the symmetric \levy\ stable distributions approximately follow power laws with probability $1/|x|^{1+\a}$ as $|x|/\g$ increases beyond a modest threshold value.

These \levy\ distributions are called ``stable'' because they have two properties that no other distributions have.  First, all infinite sums of independent observations from the same distribution converge to a \levy\ stable distribution.  Second, the properly normalized sum of two or more \levy\ stable distributions is a \levy\ stable distribution of the same form.  

These properties cause aggregations to converge to the \levy\ stable form.  Once an aggregate has converged to this form, combining with another aggregate tends to keep the form stable.  For these reasons, the \levy\ stable distributions play a dominant role in the patterns of nature.

\subsection*{Maximum entropy: information and the stable distributions}

I showed in \Eq{spectralMaxEntGauss} that 
	$$H(s) = e^{-\g^2s^2}$$
is the maximum entropy pattern in the Fourier domain given information about the spectral variance.  The spectral variance in this case is $\int s^2H'(s)\ds=\ang{s^2}$.  If we follow the same derivation for maximum entropy in the Fourier domain, but use the general expression for the $\a$th spectral moment for $\a\le2$, given by $\int |s|^\a H'(s)\ds=\ang{|s|^\a}$, we obtain the general expression for maximum entropy subject to information about the $\a$th spectral moment as
\begin{equation}\label{levyMaxEnt}
	H(s) = e^{-\g^\a |s|^\a}. 
\end{equation}
This expression matches the general form obtained by $n$-fold convolution of the sum in \Eq{levySum} converging to \Eq{levyFourier} by Fourier analysis.  The value of $\a$ does not have to be an integer: it can take on any value between 0 and 2. [Here, $\g^\a=1/\a\ang{|s|^\a}$, and, in \Eq{fourierMaxEnt}, $k=\a\g/2\Gamma(1/a)$.]

The sum in \Eq{levySum} is a particular generative model that leads to the Fourier pattern given by \Eq{levyFourier}.  By contrast, the maximum entropy model uses only information about the $\a$th spectral moment and derives the same result.  The maximum entropy analysis has no direct tie to a generative model. The maximum entropy result shows that any generative model or any other sort of information or assumption that sets the same informational constraints yields the same pattern.  

What does the $\a$th spectral moment mean?   For $\a=2$, the moment measures the variance in frequency when we weight frequencies by their intensity of contribution to pattern.  For $\a=1$, the moment measures the average frequency weighted by intensity.   In general, as $\a$ declines, we weight more strongly the lower frequencies in characterizing the distribution of intensities.  Lower frequencies correspond to more extreme values in the direct domain, $x$, because low frequencies waves spread more widely.  So, as $\a$ declines, we weight more heavily the tails of the probability distribution in the direct domain.  In fact, the weighting $\a$ corresponds exactly to the weight in the tails of $1/|x|^\a$ for large values of $x$.

Numerous papers discuss spectral moments \shortcite<e.g., >{Benoit92method,Eriksson04Imitations}.  I could find in the literature only a very brief mention of using maximum entropy to derive \Eq{levyMaxEnt} as a general expression of the symmetric \levy\ stable distributions \shortcite{Bologna02equilibrium}.  It may be that using spectral moments in maximum entropy is not considered natural by the physicists who work in this area.  Those physicists have discussed extensively alternative definitions of entropy by which one may understand the stable distributions \cite<e.g., >{Abe00theorem}.  My own view is that there is nothing unnatural about spectral moments.  The Fourier domain captures the essential features by which aggregation shapes information.

\subsection*{Truncated \levy\ flights}

The variance is not infinite in practical applications.  Finite variance means that aggregates eventually converge to a Gaussian as the number of the components in the aggregate increases.  Yet many observable patterns have the power law tails that characterize the \levy\ distributions that arise as attractors with infinite variance.  Several attempts have been made to resolve this tension between the powerful attraction of the Gaussian for finite variance and the observable power law patterns.  The issue remains open.  In this section, I make a few comments about the alternative perspectives of generative and informational views.

The generative approach often turns to truncated \levy\ flights to deal with finite variance \cite{Mantegna94flight,Mantegna95index,Voit05Markets,Mariani07indices}.  In the simplest example, each observation comes from a distribution with a power law tail such as \Eq{powerLaw} with $L=1$, repeated here
	$$p_y = \d  y^{-(1+\d)},$$
for $1\le y < \infty$.  The variance of this distribution is infinite.  If we truncate the tail such that $1\le y \le U$, and normalize so that the total probability is one, we get the distribution
	$$p_y = \ovr{\d  y^{-(1+\d)}}{1-U^{-\d}},$$
which for large $U$ is essentially the same distribution as the standard power law form, but with the infinite tail truncated.  The truncated distribution has finite variance.  Aggregation of the truncated power law will eventually converge to a Gaussian.  But the convergence takes a very large number of components, and the convergence can be very slow.  For practical cases of finite aggregation, the sum will often look somewhat like a power law or a \levy\ stable distribution.

I showed earlier that power laws arise by maximum entropy when one has information only about $\ang{\log(y)}$ and the lower bound, $L$.  Such distributions have infinite variance, which may be unrealistic.  The assumption that the variance must be finite means that we must constrain maximum entropy to account for that assumption.  Finite variance implies that $\ang{y^2}$ is finite.  We may therefore consider the most likely pattern arising simply from maximum entropy subject to the constraints of minimum value, $L$, geometric mean characterized by $\ang{\log(y)}$, and finite second moment given by $\ang{y^2}$. For this application, we will usually assume that the second moment is large to preserve the power law character over most of the range of observations.  With those assumptions, our standard procedure for maximum entropy in \Eq{genMaxEnt} yields the distribution 
\begin{equation}\label{maxEntTruncPower}
	p_y=ky^{-(1+\d)}e^{-\g y^2}, 
\end{equation}
where I have here used notation for the $\l$ constants of \Eq{genMaxEnt} with the substitutions $\l_1=1+\d$ and $\l_2=\g$. No simple expressions give the values of $k$, $\d$, and $\g$; those values can be calculated from the three constraints: $\int p_y\dy=k$, $\int y^2p_y\dy=\ang{y^2}$, and $\int \log(y)p_y\dy=\ang{\log(y)}$.  If we assume large but finite variance, then $\ang{y^2}$ is large and $\g$ will be small.  As long as values of $\g y^2$ remain much less than one, the distribution follows the standard power law form of \Eq{powerLaw}.  As $\g y^2$ grows above one, the tail of the distribution approaches zero more rapidly than a Gaussian.

I emphasize this informational approach to truncated power laws because it seems most natural to me.  If all we know is a lower bound, $L$, a power law shape of the distribution set by $\d$ through the observable range of magnitudes, and a finite variance, then the form such as \Eq{maxEntTruncPower} is most likely.

\subsection*{Why are power laws so common?}

Because spectral distributions tend to converge to their maximum entropy form
\begin{equation}\label{spectralPowerLaw}
	H(s) = e^{-\g^\a |s|^\a}.  
\end{equation}
With finite variance, $\a$ very slowly tends toward $2$, leading to the Gaussian for aggregations in which component perturbations have truncated fat tails with finite variance.  If the number of components in the aggregate is not huge, then such aggregates may often closely follow \Eq{spectralPowerLaw} or a simple power law through the commonly measurable range of values, as in \Eq{maxEntTruncPower}.  Put another way, the geometric mean often captures most of the information about a process or a set of data with respect to underlying distribution.

This informational view of power laws does not favor or rule out particular hypotheses about generative mechanisms.  For example, word usage frequencies in languages might arise by particular processes of preferential attachment, in which each additional increment of usage is allocated in proportion to current usage.  But we must recognize that any process, in the aggregate, that preserves information about the geometric mean, and tends to wash out other signals, converges to a power law form.  The consistency of a generative mechanism with observable pattern tells us little about how likely that generative mechanism was in fact the cause of the observed pattern.  Matching generative mechanism to observed pattern is particularly difficult for common maximum entropy patterns, which are often attractors consistent with many distinct generative processes. 

\section*{Extreme value theory}

The \levy\ stable distributions express widespread patterns of nature that arise through summation of perturbations.  Summing logarithms is the same as multiplication. Thus, the \levy\ stable distributions, including the special Gaussian case, also capture multiplicative interactions.

Extreme values define the other great class of stable distributions that shape the common patterns of nature.  An extreme value is the largest (or smallest) value from a sample.  I focus on largest values.  The same logic applies to smallest values.

The cumulative probability distribution function  for extreme values, $G(x)$, gives the probability that the greatest observed value in a large sample will be less than $x$.  Thus, $1-G(x)$ gives the probability that the greatest observed value in a large sample will be higher than $x$. 

Remarkably, the extreme value distribution takes one of three simple forms. The particular form depends only on the shape of the upper tail for the underlying probability distribution that governs each observation.  In this section, I give a brief overview of extreme value theory and its relation to maximum entropy.  As always, I emphasize the key concepts.  Several books present full details, mathematical development, and numerous applications \shortcite{Embrechts97Finance,Kotz00Applications,Coles01Values,Gumbel04Extremes}.

\subsection*{Applications}

Reliability, time to failure, and mortality may depend on extreme values.  Suppose an organism or a system depends on numerous components.  Failure of any component causes the system to fail or the organism to die.  One can think of failure for a component as an extreme value in a stochastic process.  Then overall failure depends on how often an extreme value arises in any of the components.  In some cases, overall failure may depend on breakdown of several components. The Weibull distribution is often used to describe these kinds of reliability and failure problems \cite{Juckett92risks}.  We will see that the Weibull is one of the three general types of extreme value distributions.

Many problems in ecology and evolution depend on evaluating rare events.  What is the risk of invasion by a pest species \shortcite{Franklin08compliance}?  For an endangered species, what is the risk of rare environmental fluctuations causing extinction?  What is the chance of a rare beneficial mutation arising in response to a strong selective pressure \shortcite{Beisel07effects}?

\subsection*{General expression of extreme value problems}

Suppose we observe a random variable, $Y$, with a cumulative distribution function, or cdf, $F(y)$. The cdf is defined as the probability that an observation, $Y$, is less than $y$.  In most of this paper, I have focused on the probability distribution function, or pdf, $f(y)$.  The two expressions are related by 
	$$P(Y<y) = F(y) = \int_{-\infty}^y f(x)\dx.$$
The value of $F(y)$ can be used to express the total probability in the lower tail below $y$.  We often want the total probability in the upper tail above $y$, which is
	$$P(Y>y) = 1-F(y)=\Fhat(y)=\int^{\infty}_y f(x)\dx,$$
where I use $\Fhat$ for the upper tail probability.

Suppose we observe $n$ independent values $Y_i$ from the distribution $F$.  Define the maximum value among those $n$ observations as $M_n$.  The probability that the maximum is less than $y$ is equal to the probability that each of the $n$ independent observations is less than $y$, thus
	$$P(M_n<y) = [F(y)]^n.$$
This expression gives the extreme value distribution, because it expresses the probability distribution of the maximum value.  The problem is that, for large $n$, $[F(y)]^n\to0$ if $F(y)<1$ and $[F(y)]^n\to1$ if $F(y)=1$.  In addition, we often do not know the particular form of $F$.  For a useful analysis, we want to know about the extreme value distribution without having to know exactly the form of $F$, and we want to normalize the distribution so that it approaches a limiting value as $n$ increases.  We encountered normalization when studying sums of random variables: without normalization, a sum of $n$ observations often grows infinitely large as $n$ increases.  By contrast, a properly normalized sum converges to a \levy\ stable distribution.

For extreme values, we need to find a normalization for the maximum value in a sample of size $n$, $M_n$, such that
\begin{equation}\label{evtNorm}
	P[(M_n - b_n)/a_n < y] \to G(y). 
\end{equation}
In words, if we normalize the maximum, $M_n$, by subtracting a location coefficient, $b_n$, and dividing by a scaling coefficient, $a_n$, then the extreme value distribution converges to $G(y)$ as $n$ increases.  Using location and scale coefficients that depend on $n$ is exactly what one does in normalizing a sum to obtain the standard Gaussian with mean zero and standard deviation one: in the Gaussian case, to normalize a sum of $n$ observations, one subtracts from the sum $b_n=n\m$, and divides the sum by $\sqrt{n}\s$, where $\m$ and $\s$ are the mean and standard deviation of the distribution from which each observation is drawn.  The concept of normalization for extreme values is the same, but the coefficients differ, because we are normalizing the maximum value of $n$ observations rather than the sum of $n$ observations.

We can rewrite our extreme value normalization in \Eq{evtNorm} as
	$$P(M_n <a_n y+b_n) \to G(y).$$
Next we use the equivalences established above to write
	$$P(M_n <a_n y+b_n)=\big[F(a_n y+b_n)\big]^n=\big[1-\Fhat(a_n y+b_n)\big]^n.$$
We note a very convenient mathematical identity
	$$\bigg(1-\ovr{\Fhat(y)}{n}\bigg)^n\to e^{-\Fhat(y)}$$
as $n$ becomes large.  Thus, if we can find values of $a_n$ and $b_n$ such that 
\begin{equation}\label{evtUpperNorm}
	\Fhat(a_n y+b_n) = \ovr{\Fhat(y)}{n}, 
\end{equation}
then we obtain the general solution for the extreme value problem as
\begin{equation}\label{evtGenSol}
	G(y) = e^{-\Fhat(y)}, 
\end{equation}
where $\Fhat(y)$ is the probability in the upper tail for the underlying distribution of the individual observations.  Thus, if we know the shape of the upper tail of $Y$, and we can normalize as in \Eq{evtUpperNorm}, we can express the distribution for extreme values, $G(y)$.

\subsection*{The tail determines the extreme value distribution}

I give three brief examples that characterize the three different types of extreme value distributions.  No other types exist \cite{Embrechts97Finance,Kotz00Applications,Coles01Values,Gumbel04Extremes}.

In the first example, suppose the upper tail of $Y$ decreases exponentially such that $\Fhat(y)=e^{-y}$.  Then, in \Eq{evtUpperNorm}, using $a_n=1$ and $b_n=-\log(n)$, from \Eq{evtGenSol}, we obtain
\begin{equation}\label{evtGumbel}
	G(y)=e^{-e^{-y}}, 
\end{equation}
which is called the double exponential or Gumbel distribution.  Typically any distribution with a tail that decays faster than a power law attracts to the Gumbel, where, by \Eq{powerLaw}, a power law has a total tail probability in its cumulative distribution function proportional to $1/y^\d$, with $\d<2$.  Exponential, Gaussian, and gamma distributions all decay exponentially with tail probabilities less than power law tail probabilities.

In the second example, let the upper tail of $Y$ decrease like a power law such that $\Fhat(y)=y^{-\d}$.  Then, with $a_n=n^{1/\d}$ and $b_n=0$, we obtain 
\begin{equation}\label{evtFrechet}
	G(y)=e^{-y^{-\d}}, 
\end{equation}
which is called the Fr{\'e}chet distribution. 

Finally, if $Y$ has a finite maximum value $M$ such that $\Fhat(y)$ has a truncated upper tail, and the tail probability near the truncation point is $\Fhat(y)=(M-y)^\d$, then, with $a_n=n^{-1/\d}$ and $b_n=n^{-1/\d}M - M$, we obtain 
\begin{equation}\label{evtWeibull}
	G(y)=e^{-(M-y)^\d}, 
\end{equation}
which is called the Weibull distribution.  Note that $G(y) = 0$ for $y>M$, because the extreme value can never be larger than the upper truncation point.

\subsection*{Maximum entropy: what information determines extreme values?}

In this section, I show the constraints that define the maximum entropy patterns for extreme values.  Each pattern arises from two constraints.  One constraint sets the average location either by the mean, $\ang{y}$, for cases with exponential tail decay, or by the geometric mean measured by $\ang{\log(y)}$ for power law tails.  To obtain a general form, express this first constraint as $\ang{\xi(y)}$, where $\xi(y)=y$ or $\xi(y)=\log(y)$. The other constraint measures the average tail weighting, $\ang{\Fhat(y)}$.

With the two constraints $\ang{\xi(y)}$ and $\ang{\Fhat(y)}$, the maximum entropy probability distribution (pdf) is
\begin{equation}\label{evtMaxEntGen}
	g(y) = ke^{-\l_1\xi(y)-\l_2\Fhat(y)}. 
\end{equation}
We can relate this maximum entropy probability distribution function (pdf) to the results for the three types of extreme value distributions. I gave the extreme value distributions as $G(y)$, the cumulative distribution function (cdf).  We can obtain the pdf from the cdf by differentiation, because $g(y)=\dd G(y)/\dy$.  

From \Eq{evtGumbel}, the pdf of the Gumbel distribution is
	$$g(y) = e^{-y-e^{-y}},$$
which, from \Eq{evtMaxEntGen}, corresponds to a constraint $\ang{\xi(y)}=\ang{y}$ for the mean and a constraint $\ang{\Fhat(y)}=\ang{e^{-y}}$ for the exponentially weighted tail shape. Here, $k=\l_1=\l_2=1$.

From \Eq{evtFrechet}, the pdf of the Fr{\'e}chet distribution is
	$$g(y) = \d y^{-(1+\d)}e^{-y^{-\d}},$$
which, from \Eq{evtMaxEntGen}, corresponds to a constraint $\ang{\xi(y)}=\ang{\log(y)}$ for the geometric mean and a constraint $\ang{\Fhat(y)}=\ang{y^{-\d}}$ for power law weighted tail shape. Here, $k=\d$, $\l_1=1+\d$, and $\l_2=1$.

From \Eq{evtWeibull}, the pdf of the Weibull distribution is
	$$g(y) = \d (M-y)^{\d-1}e^{-(M-y)^{\d}},$$
which, from \Eq{evtMaxEntGen}, corresponds to a constraint $\ang{\xi(y)}=\ang{\log(y)}$ for the geometric mean and a constraint $\ang{\Fhat(y)}=\ang{(M-y)^{\d}}$ that weights extreme values by a truncated tail form. Here, $k=\d$, $\l_1=1-\d$, and $\l_2=1$.

In summary, \Eq{evtMaxEntGen} provides a general form for extreme value distributions.  As always, we can think of that maximum entropy form in two complementary ways.  First, aggregation by repeated sampling suppresses weak signals and enhances strong signals until the only remaining information is contained in the location and in the weighting function constraints.  Second, independent of any generative process, if we use measurements or extrinsic information to estimate or assume location and weighting function constraints, then the most likely distribution given those constraints takes on the general extreme value form.

\section*{Generative models versus information constraints}

The derivations in the previous section followed a generative model in which one obtains $n$ independent observations from the same underlying distribution.  As $n$ increases, the extreme value distributions converge to one of three forms, the particular form depending on the tail of the underlying distribution.

We have seen several times in this paper that such generative models often attract to very general maximum entropy distributions.  Those maximum entropy distributions also tend to attract a wide variety of other generative processes. In the extreme value problem, any underlying distributions that share similar probability weightings in the tails fall within the domain of attraction to one of the three maximum entropy extreme value distributions \cite{Embrechts97Finance}.  

In practice, one often first discovers a common and important pattern by a simple generative model.  That generative model aggregates observations drawn independently from a simple underlying distribution that may be regarded as purely random or neutral.  It is, however, a mistake to equate the neutral generative model with the maximum entropy pattern that it creates.  Maximum entropy patterns typically attract a very wide domain of generative processes.  The attraction to simple maximum entropy patterns arises because those patterns express simple informational constraints and nothing more.  Aggregation inevitably washes out most information by the accumulation of partially uncorrelated perturbations.  What remains in any aggregation is the information in the sampling structure, the invariance to changes in measurement scale, and the few signals retained through aggregation.  Those few bits of information define the common patterns of nature.  

Put another way, the simple generative models can be thought of as tools by which we discover important maximum entropy attractor distributions.  Once we have found such distributions by a generative model, we may extract the informational constraints that define the pattern.  With that generalization in hand, we can then consider the broad scope of alternative generative processes that preserve the information that defines the pattern.  The original generative model no longer has special status---our greatest insight resides with the informational constraints that define the maximum entropy distribution.

The challenge concerns how to use knowledge of the common patterns to draw inferences about pattern and process in biology.  This paper has been about the first step: to understand clearly how information defines the relations between generative models of process and the consequences for pattern.  I only gave the logical structure rather than direct analyses of important biological patterns.  The next step requires analysis of the common patterns in biology with respect to sampling structure, informational constraints, and the domains of attraction for generative models to particular patterns.  What range of non-neutral generative models attract to the common patterns, because the extra non-neutral information gets washed out with aggregation?  

With the common patterns of biology better understood, one can then analyze departures from the common patterns more rationally.  What extra information causes departures from the common patterns?  Where does the extra information come from?  What is it about the form of aggregation that preserves the extra information?  How are evolutionary dynamics and biological design influenced by the tendency for aggregates to converge to a few common patterns?  

\section*{Acknowledgments}

I came across the epigraph from Gnedenko and Kolmogorov in \citeA{Sornette06Tools} and the epigraph from Galton in \citeA{Hartl00genetics}. My research is supported by National Science Foundation grant EF-0822399, National Institute of General Medical Sciences MIDAS Program grant U01-GM-76499, and a grant from the James S.~McDonnell Foundation.  

\vfill\eject

\bibliography{Biblio/common1}
\bibliographystyle{latexStyles/apaciteJEB}

\end{document}